\RequirePackage{fix-cm}
\documentclass[smallextended]{svjour3}       
\smartqed  
\usepackage{appendix}
\usepackage{amsmath}
\usepackage{graphicx}
\usepackage{lineno}
\usepackage{array}
\usepackage{longtable}
\usepackage{natbib}
\usepackage{ulem}
\usepackage{tikz}
\usetikzlibrary{shapes.geometric}
\usepackage{pgfplots}
\pgfplotsset{compat=newest}

\newcommand{\greycross}{\raisebox{0pt}{\tikz{\draw[gray, line width=1.0pt] (0,0) -- (2mm,2mm) (0,2mm) -- (2mm,0);}}}
\newcommand{\blackbordergrayfilledcircle}{\raisebox{0pt}{
  \tikz{\fill[gray] (0, 0) circle (2.5pt);
        \draw[black, very thick] (0, 0) circle (2.5pt);}}}
\newcommand{\greyedgediamond}{\raisebox{0pt}{%
  \tikz{\draw[gray, thick] (0,0) -- (1mm,1mm) -- (2mm,0) -- (1mm,-1mm) -- cycle;}}}
\newcommand{\greyedgedsquare}{\raisebox{0pt}{%
  \tikz{\draw[gray, thick] (0,0) rectangle (2mm,2mm);}}}
\newcommand{\greyplus}{\raisebox{0pt}{\tikz{\draw[gray, line width=1.0pt] (0,1mm) -- (2mm,1mm) (1mm,0) -- (1mm,2mm);}}}
\newcommand{\filledgreyedgediamond}{\raisebox{0pt}{%
  \tikz{\fill[gray] (0,0) -- (1mm,1mm) -- (2mm,0) -- (1mm,-1mm) -- cycle;
        \draw[black, thick] (0,0) -- (1mm,1mm) -- (2mm,0) -- (1mm,-1mm) --cycle;}}}
\newcommand{\greyedgedcircle}{\raisebox{0pt}{%
  \tikz{\draw[gray, thick] (1mm,1mm) circle (1mm);}}}
\pgfdeclareplotmark{mystar}{
    \node[star,star point ratio=2.25,minimum size=6pt,
          inner sep=0pt,draw=black,solid,fill=gray] {};}
\newcommand{\filledgreypentagram}{\raisebox{0pt}{%
  \tikz{\pgfuseplotmark{mystar};}}}
\newcommand{\greyfilledblackedgetriangledownward}{\raisebox{0pt}{%
\tikz{\fill[gray] (0,0) -- (1.5mm,0) -- (0.75mm,-1.5mm) -- cycle;
        \draw[black, thick] (0,0) -- (1.5mm,0) -- (0.75mm,-1.5mm) -- cycle;}}}
\newcommand{\greyfilledblackedgetriangle}{\raisebox{0pt}{%
\tikz{\fill[gray] (0,0) -- (1.5mm,0) -- (0.75mm,1.5mm) -- cycle;
        \draw[black, thick] (0,0) -- (1.5mm,0) -- (0.75mm,1.5mm) -- cycle;}}}
\newcommand{\blackfilledcircle}{\raisebox{0pt}{
\tikz{\fill[black] (0, 0) circle (2.5pt);}}}
\newcommand{\redsquarefilled}{\raisebox{0pt}{\tikz{\fill[red] (0,0) rectangle (2mm,2mm);}}}
\newcommand{\reddiamondfilled}{\raisebox{0pt}{\tikz{\fill[red] (0,0) -- (1mm,1mm) -- (2mm,0) -- (1mm,-1mm) -- cycle;}}}
\newcommand{\greensquarefilled}{\raisebox{0pt}{\tikz{\fill[green] (0,0) rectangle (2mm,2mm);}}}
\newcommand{\greendiamondfilled}{\raisebox{0pt}{\tikz{\fill[green] (0,0) -- (1mm,1mm) -- (2mm,0) -- (1mm,-1mm) -- cycle;}}}
\newcommand{\magentafilledcircle}{\raisebox{0pt}{
\tikz{\fill[magenta] (0, 0) circle (2.5pt);}}}
\newcommand{\bluesquarefilled}{\raisebox{0pt}{\tikz{\fill[blue] (0,0) rectangle (2mm,2mm);}}}
\newcommand{\bluediamondfilled}{\raisebox{0pt}{\tikz{\fill[blue] (0,0) -- (1mm,1mm) -- (2mm,0) -- (1mm,-1mm) -- cycle;}}}
\newcommand{\bluetrianglefilled}{\raisebox{0pt}{%
  \tikz{\fill[blue] (0,0) -- (2mm,0) -- (1mm,2mm) -- cycle;}}}
\newcommand{\bluetriangledownfilled}{\raisebox{0pt}{%
  \tikz{\fill[blue] (0,0) -- (2mm,0) -- (1mm,-2mm) -- cycle;}}}
\newcommand{\bluetrianglerightfilled}{\raisebox{0pt}{%
  \tikz{\fill[blue] (0,0) -- (2mm,1mm) -- (0,2mm) -- cycle;}}}

\newcommand{\greyfilledblackedgesquare}{\raisebox{0pt}{%
  \tikz{\fill[gray] (0,0) rectangle (2mm,2mm);
        \draw[black, thick] (0,0) rectangle (2mm,2mm);}}}

\setcitestyle{aysep={}}
\newcommand*\patchAmsMathEnvironmentForLineno[1]{%
\expandafter\let\csname old#1\expandafter\endcsname\csname #1\endcsname
\expandafter\let\csname oldend#1\expandafter\endcsname\csname end#1\endcsname
\renewenvironment{#1}%
{\linenomath\csname old#1\endcsname}%
{\csname oldend#1\endcsname\endlinenomath}}% 
\newcommand*\patchBothAmsMathEnvironmentsForLineno[1]{%
\patchAmsMathEnvironmentForLineno{#1}%
\patchAmsMathEnvironmentForLineno{#1*}}%
\AtBeginDocument{%
\patchBothAmsMathEnvironmentsForLineno{equation}%
\patchBothAmsMathEnvironmentsForLineno{align}%
\patchBothAmsMathEnvironmentsForLineno{flalign}%
\patchBothAmsMathEnvironmentsForLineno{alignat}%
\patchBothAmsMathEnvironmentsForLineno{gather}%
\patchBothAmsMathEnvironmentsForLineno{multline}%
}

\begin{document}

\title{A Moving Surface Drag Model for LES of Wind over Waves
}

\author{Manuel Ayala \and   
  Zein Sadek \and Ondřej Ferčák\and 
  Raul Bayo\'{a}n Cal\and 
  Dennice F. Gayme\and 
  Charles Meneveau
}

\institute{Manuel Ayala  \at
              Mechanical Engineering, Johns Hopkins University, Baltimore, MD 21218-2625, USA \\
              \email{mayala5@jhu.edu}           
           \and
           Zein Sadek \at
             Mechanical and Materials Engineering, Portland State University, Portland, OR 97201, USA
            \and
            Ondřej Ferčák \at
            Mechanical and Materials Engineering, Portland State University, Portland, OR 97201, USA
            \and
            Raul Bayo\'{a}n Cal \at
            Mechanical and Materials Engineering, Portland State University, Portland, OR 97201, USA
            \and
            Dennice F. Gayme \at
            Mechanical Engineering, Johns Hopkins University, Baltimore, MD 21218-2625, USA
            \and
            Charles Meneveau \at
            Mechanical Engineering, Johns Hopkins University, Baltimore, MD 21218-2625, USA
}

\date{Received: DD Month YEAR / Accepted: DD Month YEAR}

\maketitle

\begin{abstract}
Numerical prediction of the interactions between wind and ocean waves is essential for climate modeling and a wide range of offshore operations. Large Eddy Simulation (LES) of the marine atmospheric boundary layer is a practical numerical predictive tool but requires parameterization of surface fluxes at the air-water interface. Current momentum flux parameterizations primarily use wave-phase adapting computational grids, incurring high computational costs, or use an equilibrium model based on Monin-Obukhov similarity theory for rough surfaces that cannot resolve wave phase information. To include wave phase-resolving physics at a cost similar to the equilibrium model, the MOving Surface Drag (MOSD) model is introduced.  
It assumes ideal airflow over locally piece-wise planar representations of moving water wave surfaces.
Horizontally unresolved interactions are still modeled using the equilibrium model. Validation against experimental and numerical datasets with known monochromatic waves demonstrates the robustness and accuracy of the model in representing wave-induced impacts on mean velocity and Reynolds stress profiles. The model is formulated to be applicable to a broad range of wave fields and its ability to represent cross-swell and multiple wavelength cases is illustrated. Additionally, the model is applied to LES of a laboratory-scale fixed-bottom offshore wind turbine model, and the results are compared with wind tunnel experimental data. The LES with the MOSD model shows good agreement in wind-wave-wake interactions and phase-dependent physics at a low computational cost. The model's simplicity and minimal computational needs make it valuable for studying turbulent atmospheric-scale flows over the sea, particularly in offshore wind energy research.

\end{abstract}

\section{Introduction}
\label{intro}
The interaction between wind and waves is a complicated multiscale process that occurs at the interface between the atmosphere and the ocean. Understanding the dynamics of wind-wave interactions is crucial not only for advancing our fundamental knowledge of the underlying physics, but also for practical applications such as weather forecasting, coastal engineering, and marine operations, such as offshore wind farm design and control. 
Extensive research has been devoted to investigating the wave-induced effects on the vertical structure of turbulent airflow above the ocean or wavy surfaces, with valuable insights gained from field measurements \citep{grachev2001upward,donelan2006,edson2007,grare2018} and laboratory experiments \citep{snyder1981,banner_peirson_1998,Funke2021}. These studies have facilitated the quantification of momentum transfer between the sea and the air medium. More recently, \cite{buckley2016}, \cite{buckley2020} and \cite{yousefi_2020} conducted meticulous particle image velocimetry (PIV) measurements of turbulent air above wind-generated waves, shedding light on the intricate airflow structure and wind stress above the wave surface. Their work quantified the impact of form (pressure) drag on the total air-water momentum flux,  comparing it to the effects of viscous drag and its dependence on wave slope. Experimental studies by \cite{peirson_garcia_2008} and \cite{Grare_2013} enabled the estimation of the wave growth rate and dissipation rate due to the energy transfer between wind and wave fields.

In recent years, numerical simulations have emerged as a vital tool for investigating the fundamental physics of turbulent flows. Early numerical models of turbulent flow over wavy surfaces \citep{gent_taylor_1976,al-zanaidi_hui_1984,li2000numerical} relied on Reynolds-averaged equations and eddy viscosity formulations, which yielded diverse predictions of wave growth due to the sensitivity of these quantities to the chosen turbulence closure scheme.  Direct Numerical Simulations (DNS) have been employed as a much higher fidelity numerical approach to examine turbulent flows over complex geometries, such as wavy surfaces \citep{sullivan_2000,yang_shen_2010,Deskos2022,wu_popinet_deike_2022}. While DNS has no need for parameterizations or subgrid-scale modeling, it is inherently limited to low or moderate Reynolds numbers due to the substantial computational costs associated with resolving viscous sublayer processes. In contrast to DNS, Large Eddy simulation (LES) offers the advantage of lower cost since   the small scales of turbulence are modeled via a subgrid-scale model while still resolving the large-scale motions. 
Interactions between the atmospheric boundary layer (ABL) and waves have been extensively investigated using wall-resolved LES (WRLES), in which a no-slip boundary condition is applied to the wavy surface using sufficiently fine spatial resolution near the surface.  Several WRLES studies study monochromatic waves (sinusoidal wave train) as the wavy boundary \citep{sullivan2008large,aakervik2019role,Zhang2019,Wang_2021,cao_shen_2021,Hao2021}, while others consider a complete ocean wave spectrum \citep{yang_wavespec_2011,hao_specwave_2019,wang_zhang_hao_huang_shen_xu_zhang_2020}.  The viscous sublayer in the marine atmospheric boundary layer (MABL) is typically only fractions of millimeter thick. Therefore, WRLES is  still limited to  moderate Reynolds numbers, since its need to resolve near-wall structures makes its computational cost scaling not significantly lower than that of DNS  \citep{XYang_2021}. For both WRLES and DNS of turbulent flow over a waves, a boundary fitted grid or a terrain-following grid is typically employed to represent the surface in the numerical domain \citep{Deskos2021}, often called ``wave-phase-resolved'' approach. The use of this approach coupled with the high resolution near the wall enables an accurate resolution of phase-dependent wave effects and near-wall wave-induced turbulent effects. However, this comes at the expense of high and frequently prohibitive computational costs.

More cost-effective numerical simulations of wind fields can be accomplished using wall-modeled LES (WMLES), where the near-wall region must be modelled through wall models. Over the past decades a number of wall models have been developed  for flow over on-shore rough terrain \citep{stoll2020large}, typically based on ``Monin-Obukhov Similarity Theory'' (MOST) \citep{moeng1984} in geophysical applications or the logarithmic law in engineering applications \citep{piomelliBalaras}. Within the MOST framework, the surface stress is determined by a surface roughness parameter ($z_o$) that incorporates the influence of unresolved surface features on the flow. However, the modeling of flow over moving surface waves has received less attention. The classical approach, known as the ``wave-phase-averaged'' approach, involves specifying an effective roughness length specifically for ocean waves. The Charnock model, introduced by \cite{charnock},  is a widely adopted approach for parameterizing an equivalent effective roughness for ABLs over ocean waves. In  recent years, several investigations have proposed other surface roughness relationships based on a wave-age and wave-steepness parametrization \citep{masuda,toba,smith}.  However, all of the ``wave-phase-averaged'' approaches are highly dependent on empirical parameters, which must be tuned and can vary significantly from dataset to dataset \citep{Deskos2022}. Moreover, such methods cannot represent phase-dependent processes. 

To address the current limitations of wind-wave interaction models, we aim to  develop an approach that can resolve some phase-dependent wave effects while preserving the computational efficiency offered by phase-averaged approaches. Recently, \cite{aditya} have introduced a
generalization of the horizontally resolved, vertically unresolved surface gradient model of \cite{AndersonMeneveau2010} that is based on momentum fluxes associated with the velocity differences between incoming wind and wave components. For applications to wave surfaces consisting of multiple wavelengths, the model by \cite{aditya} uses a superposition over the various constitutive wavenumbers of the imposed wave spectrum and introduces a wavelength-dependent drag coefficient.  This prefactor can recover theoretical predictions of wave energy input rates at various wavelengths. With similar motivations, in this work we introduce a model that is not based on a superposition of forces over wave frequencies or wavelengths, but instead is based directly on knowledge of the local surface slope and its time evolution in physical space, without recourse to spectral descriptions and wavelength-dependent drag coefficients. The proposed MOving Surface Drag (MOSD) model  relies on the assumption that windward facing portions of the wave are exposed to increased pressure distribution. This pressure field is modeled as that arising in ideal potential flow over an inclined plane (ramp flow), while the force from the pressure field on the back leeward side of the wave is neglected altogether based on the assumption of flow separation or near separation. Drag caused by horizontally unresolved scales (small-scale interactions) are still modeled using the traditional equilibrium wall model. The MOSD model enables us to represent phase-dependent effects of ocean waves in LES using a simple flat bottom surface boundary condition at a moderate computational cost that is of similar order of magnitude to that of the equilibrium wall model.
 The main objective of the present study is to propose and validate an accurate cost-effective model for predicting the momentum transfer to the air due to sea interactions for WMLES. First, a detailed derivation and description of the MOSD model is provided in Sect. \ref{sec:MOSD_model}. In Sect. \ref{sec:les_num}   the numerical code, numerical details and suite of test benchmark cases to be used for model validation are presented. The WMLES results are compared to the benchmark cases in Sect. \ref{sec:validation}. The wide applicability of the MOSD  model is illustrated through application to more general wave conditions and propagation directions in Sect. \ref{sec:generalapps}.   The ability to introduce phase-resolved wave effects to numerical studies of offshore wind energy applications, while maintaining a low computational cost (i.e. not recurring to wave-following grids or two-fluid methods) was identified as a current challenge \citep{Deskos2021}. We showcase the applicability of the model for this application in Sect. \ref{sec:vwind_turbine_sim}, which describes LES with the MOSD air-sea  parameterization  applied to model the flow in a laboratory experiment of a fixed-bottom offshore wind turbine \citep{fercak_2022} This WMLES data are compared with the laboratory experimental data.  Finally, the main conclusions are summarized in Sect. \ref{sec:conclusions}, including a summary of the strengths and limitations of the proposed model.

\section{Moving Surface Drag (MOSD) Model}
\label{sec:MOSD_model}

The proposed MOving Surface Drag (MOSD) wall model is comprised of two components,  a model based on ideal potential flow theory applied to windward facing portions of the surface to represent form drag caused by flow blockage due to the horizontally resolved features of moving waves, and an equilibrium wall stress that follows the local law-of-the-wall (or MOST) to model effects of entirely unresolved surface roughness features.

\begin{figure}
  \centerline{\includegraphics[width=11.9cm]{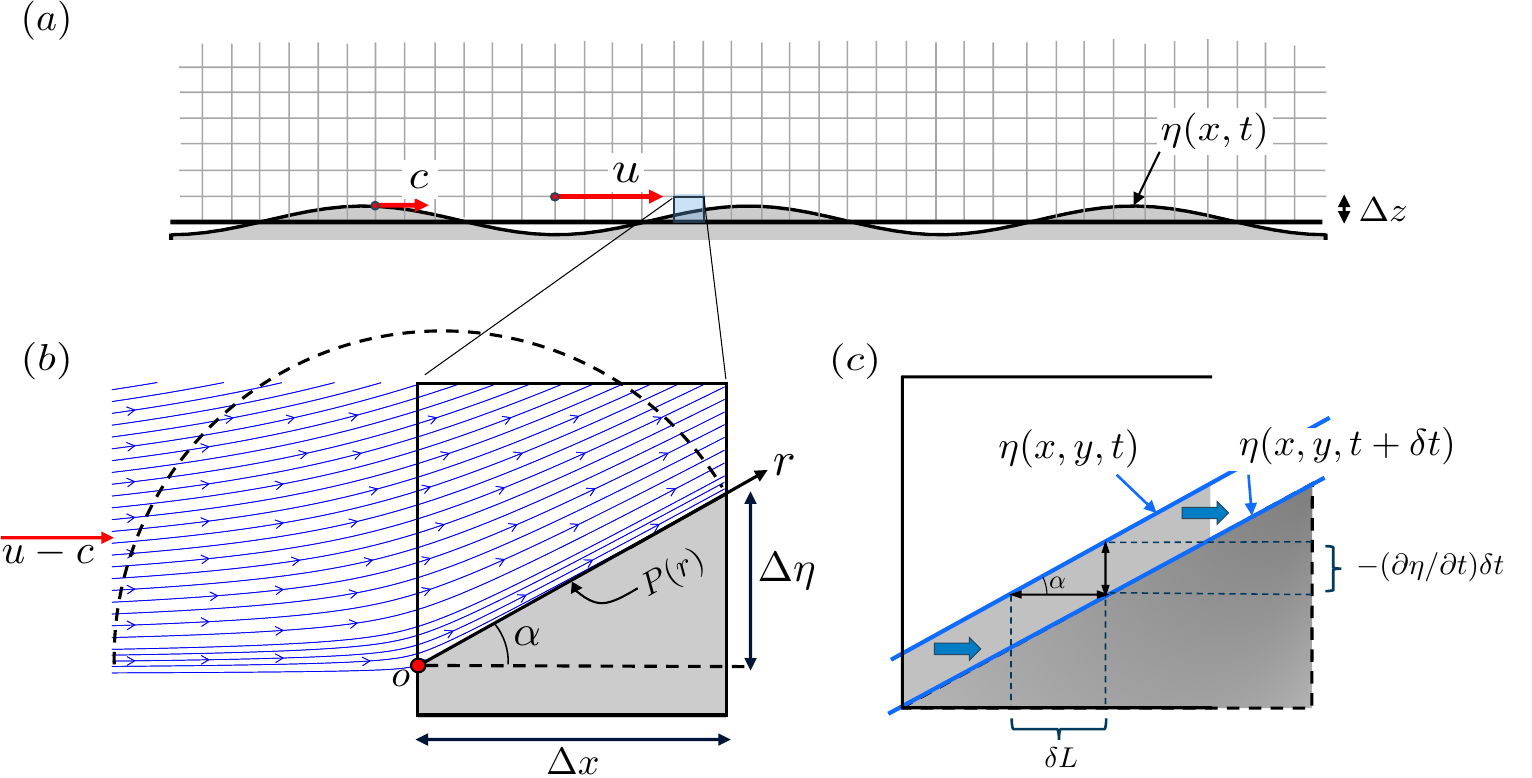}}
  \caption{(a) Schematic  of a horizontally resolved and vertically unresolved wave and the LES grid near the bottom surface. The enlargement (b) shows a segment of the surface idealized as a ramp of angle $\alpha$ and the streamlines of the assumed ideal potential flow over such a ramp used to compute the pressure distribution over the surface $P(r)$. The streamlines are based  the velocity components of the potential solution and are shown to illustrate the flow deflection caused by the ramp. The undisturbed reference velocity ($u-c$) is taken at a distance $\Delta x$ from the origin, shown by the dashed semi-circle.  (c) shows how the generalized wave propagation velocity ${\bf C}$ can be determined based on spatial and temporal gradient of the elevation field $\eta(x,\,y,\,t)$ that is assumed to be known}
\label{fig:potwave}
\end{figure}

\subsection{Windward Potential Flow Model}
\label{sec:pfb_model}
\cite{Lamb} shows that an inviscid and irrotational assumption (ideal flow) can be applied to represent the dynamics of air flow over monochromatic waves of small amplitude. However, this representation results in the pressure distribution along the air-sea interface being out of phase with the gradient of the surface distribution, leading to zero average stress on the interface. Thus, no net form drag can be imparted onto the air in pure potential flow. Alternatively, in the proposed model we  assume that potential flow only applies to the windward face of waves that are horizontally resolved in the LES and neglect any force coming from the leeward (downwind) portion of the surface by assuming the flow there to be separating or almost separating (the validity of this assumption will be discussed later on). In addition  in each computational cell of length $\Delta x$, the wave is represented locally as an inclined plane with inclination angle $\alpha$. The sketch shown in Fig.\ref{fig:potwave} illustrates the basic idea. For reference, we use a Cartesian frame, with $x, y, z$ being the streamwise, vertical,
and spanwise coordinates (also denoted as $x_1, x_2$, and  $x_3$ for the index notation).  Assuming a reference frame that is moving with the local horizontal velocity of the wave surface $c$, in each individual cell, we assume that the flow arrives horizontally at a relative velocity of $u-c$ at a distance $\Delta x$ from the moving surface. Importantly, $u$ is not the fluid velocity in the immediate vicinity of the surface but is similar to the velocity at the model grid point away from the surface (often denoted as $U_{\rm LES}$) which, as is usual in WMLES, can be much larger that the surface velocity. Certainly we typically expect it to be larger than $c$ i.e. that the LES grid is coarse and thus cannot resolve the critical layer.  Even if the true velocity then displays near-surface reduction of velocity towards the orbital velocity (caused by turbulent and viscous mixing), for the purpose of modeling the pressure distribution, a potential flow is assumed. This potential flow does not include a friction-induced near-surface velocity reduction (shearing). The potential flow assumption to compute the pressure distribution is justified since the vertical pressure distribution in boundary layers is typically not strongly affected by this shearing but mainly by inviscid processes depending on surface geometry. Note that we incorporate the effects of shearing between $u$ and the wave orbital velocities using the Equilibrium Wall Model (see Section \ref{sec:eqwm}).

The flow is locally deflected upward at an angle $\alpha$. The flow is assumed to be ideal potential flow over an inclined plane (ramp flow) that has a stagnation point at each left-most edge of the ramp and reaches a speed $u-c$ at its peak. In reality, the $i^{\rm th}$ ramp section of the discretized surface will have incoming flow coming from the deflected flow of the upstream $(i-1)^{\rm th}$ ramp section. However, here we assume that for each ramp section the flow arrives horizontally with a velocity $u-c$ far from the local stagnation point. Formulating a more accurate potential flow solution over the surface would ultimately require solving non-local Partial Differential Equations, which our model aims to avoid. Our contention is that the WPM approach is the most accurate purely local representation of the inviscid dynamics that dominate the pressure forces involved.  For such a flow, the streamfunction in cylindrical coordinates with origin at the lower point in the cell (where the flow is assumed to impinge against the surface)  is given by:
\begin{equation}
 \psi (r,\theta) = A \,r^n \sin{[n(\theta-\alpha)]},
\end{equation}
where $n = {\pi}/(\pi-\alpha)$, $A$ is a constant, $\alpha$ is the ramp angle and $\theta$ is the polar coordinate angle, as shown in Fig. \ref{fig:potwave}.

 On the inclined surface, the flow is purely radial and thus no velocity normal to the surface ($V_{\theta}$ = 0) exists, and the parameter $A$ can be prescribed such that the radial velocity of the potential flow, $V_r= - r^{-1} \, \partial \psi / \partial \theta$ at $\theta = \alpha$, equals $(u-c)$ far from the impingement point, i.e. when $r=\Delta x$ (small angles are assumed throughout). With this choice, the radial velocity becomes
\begin{equation}
V_r = (u-c) \biggl(\frac{r}{\Delta x}\biggr)^{n-1}.
\end{equation}

Our aim is to obtain the total pressure force acting on the surface, whose horizontal projection  represents the pressure drag due to the flow going over the ramp surface, neglecting unsteadiness effects arising from possible deformations of the wave surface.

 By choosing a moving reference frame that follows the wave, for the case of monochromatic waves, we can treat it as stationary within this frame of reference. For more general cases with superposition of different wavelengths moving at different wave speeds, the resulting wave surface deformation will in principle cause unsteadiness in the assumed potential flow (the angle $\alpha$ may be a function of time even when moving at the local speed of the surface). In the present version of the model, the effects of this unsteadiness on the pressure will be neglected. Examination of possible effects of flow unsteadiness arising from this approximation is left for future work. 
Consequently, in the present model we assume a steady potential flow and apply the steady Bernoulli equation between a distant point—where the velocity magnitude is \( |u-c| \)—and a point along the ramp. This approach allows us to determine the pressure distribution along the ramp as follows: 
\begin{equation}
\frac{1}{\rho} \Delta P(r) = \frac{1}{2}\,(u-c)^2\, \biggl[1 - \biggl(\frac{r}{\Delta x}\biggr)^{{2\alpha}/{(\pi - \alpha)}}  \biggr].
\end{equation}
The pressure vanishes at $r=\Delta x$, where the velocity of the ramp increases back to $u-c$. Similarly, the stagnation point is at the lowest point of the ramp, $r = 0$. Now the pressure distribution can be averaged spatially along the ramp to obtain:

\begin{equation}
\frac{1}{\rho} \Delta \overline{P}  = 
    \frac{1}{2} \,(u-c)^2 \,\frac{1}{\Delta x} \int_0^{\Delta x} \biggl[1 -
    \biggl(\frac{r}{\Delta x}\biggr)^{ {2\alpha}/{(\pi 
    - \alpha)}}  \biggr]dr
 =   (u-c)^2 \, \frac{\alpha}{\pi + \alpha} ,
     \label{inteqpt}
\end{equation}
where the overline ($\overline{\cdot}$) denotes a spatial average. Since Eq. \ref{inteqpt} depends on a single wavespeed $c$, it is derived with a monochromatic wave (moving in the $x$ direction) in mind.  For a wave moving in an arbitrary direction with respect to the horizontal coordinates   $(x,y)$,  the average pressure (also assuming that $\alpha$ is small, and $\alpha << \pi$) can be written as
\begin{equation}
\frac{1}{\rho} \Delta \overline{P}  = \frac{1}{\pi}\,|(\boldsymbol{u-c}) \cdot \boldsymbol{\hat{n}}|^2 \, |\boldsymbol \nabla \eta |,
\label{inteqptsimple}
\end{equation}
where we have used $\alpha \approx |\boldsymbol \nabla \eta |$ based on the small angle assumption, and $\boldsymbol \nabla \eta = (\partial \eta/\partial x,\partial \eta/\partial y)$ represents the spatial horizontal gradient vector of the surface elevation. Note, that we have projected the relative velocity vector $\boldsymbol{u-c}$ with the inward-pointing unit normal vector of the surface ($\hat{\boldsymbol n} =  \boldsymbol \nabla \eta /|\boldsymbol{\nabla} \eta|$) to ensure that only the velocity impinging normal to the surface is accounted for.

The derived mean pressure on the ramp surface  is based on the assumption of a surface moving at a given phase velocity ${\bf c}$. However, e.g., for a broadband ocean wave spectrum, the wave field  will encompass many different phase speeds, resulting in a need to consider the effects of the various wave component on the total force. The approach followed previously in \cite{aditya} was to superpose forces from each wave component each expressed as function of its wave speed. Here instead we assume that the wave superposition gives rise to the known and imposed time and position-dependent height distribution $\eta(x,y,t)$ whose local slope and local horizontal translation velocity relative to the air velocity gives rise to the local pressure drag force. 
We aim to determine the horizontal speed $C_i$ of the moving surface that consists of the geometric superposition of the various wave components. 
In the horizontal plane, the direction of propagation will be normal to the surface itself, i.e.,
\begin{equation}
C_i = \zeta \, \frac{\partial \eta}{\partial x_i}, \qquad i = 1,2
\end{equation}
where, $\zeta$ is some factor to be determined and $x_1$ and $x_2$ correspond to the $x$ and $y$ directions. During a time $\delta t$ the surface moves an amount $\delta x $ and $\delta y$ in each of the horizontal directions, or a distance $\delta L$ normal to the surface (see panel (c) of Fig. \ref{fig:potwave}). 
Since we aim to find only the horizontal component of the surface's displacement, we impose the condition that $ \delta \eta =  0 = \hat{\bf n} \cdot \boldsymbol \nabla \eta  \,\, \delta L \, + \, (\partial \eta/\partial t)\, \delta t = (\partial \eta /\partial x_i) \, \delta x_i \, + \, (\partial \eta/\partial t)\, \delta t $. Dividing by $\delta t$ yields $\partial \eta/\partial t = - (\partial \eta/\partial x_i)\,  (\delta x_i/\delta t)$. The quantity 
$\delta x_i/\delta t (=\hat{n}_i (\delta L/\delta t))$ is in fact the desired velocity $C_i$, thus 
$ \partial \eta/\partial t = - (\partial \eta/\partial x_i)\, C_i =  - \zeta (\partial \eta/\partial x_i)^2$ from which we conclude that $ \zeta= - (\partial \eta/\partial t)  / |\boldsymbol \nabla \eta |^2 $. Therefore, the local surface horizontal velocity to be used in the model is given by:
\begin{equation}
C_i = - \frac{\partial \eta}{\partial t} \frac{\partial \eta}{\partial x_i} \frac{1}{|\boldsymbol \nabla \eta |^2} \qquad i = 1,2.
\label{gen_vel}
\end{equation}
One can confirm that for the case of a simple monochromatic wave  moving in the $x$-direction, with a height distribution of $\eta = a\cos{(kx - \omega t)}$ we obtain that the general velocity is simply the phase velocity   ($C_1 = c = \omega/k$). However, Eq. \ref{gen_vel} is in principle applicable for any temporally deforming surface, such as a broadband wave field. 

Next, we evaluate the force and horizontal equivalent stress given the mean pressure distribution acting on the piece-wise linear surface whose horizontal projection has area 
$\Delta x \Delta y$, the horizontal LES mesh area. The local horizontal force due to a pressure distribution acting on the fluid is given by $\boldsymbol{F} = - \int_A{ \Delta P \, \boldsymbol{\hat{n}}}\: dA = - \Delta \overline{P} \, \boldsymbol{\hat{n}} \,\Delta A$, where the inward-pointing unit normal vector of the surface (projected onto the horizontal plane) is $\hat{\boldsymbol n} =  \boldsymbol \nabla \eta /|\boldsymbol{\nabla} \eta|$. Also, $\Delta A$ is the projection of the  surface onto a vertical plane perpendicular to the surface gradient, with height $\Delta \eta = (\partial \eta/\partial x_i) \,\Delta x_i \,\,(i=1,2)$. We can also use a horizontal square surface locally aligned with the iso-$\eta$ lines, of area $(\Delta L)^2 = (\Delta x \Delta y)$. Then $\Delta \eta = |\boldsymbol{\nabla} \eta| \, \Delta L$.
Therefore, the local horizontal force (per unit mass) over a horizontal plane of area $(\Delta L)^2$ due to the pressure force acting on $\Delta A = \Delta \eta \Delta L$ is given by:
\begin{equation}
f_i = - \frac{1}{\pi} \, |(\boldsymbol{u-C}) \cdot \boldsymbol{\hat{n}}|^2 \, |\boldsymbol{\nabla} \eta| \, \hat{n}_i  \, (|\boldsymbol{\nabla} \eta| \,\Delta L) \,\, \Delta L \qquad i = 1,2.
\end{equation}
Note that this expression has been divided by density for inclusion in the momentum equation, so we do not include density in this expression. 
Furthermore, the force can be represented by a local wall stress defined as the horizontal force per unit area, namely  $\tau_{i3} = - f_i / (\Delta L)^2$, leading to:
\begin{equation}
\tau_{i3} = \frac{1}{\pi} |(\boldsymbol{u-C}) \cdot \boldsymbol{\hat{n}}|^2|\boldsymbol{\nabla} \eta|^2 \, \hat{n}_i. 
\end{equation} 
Next we postulate that we only need to account for the pressure on the windward side. On the leeside portion we assume that a displacement of the streamlines causes a pressure drop, such that the pressure force there can be neglected. This is a strong assumption of the model and may not hold in all cases. However, experiments by \cite{buckley2020} and \cite{veron2007}  have shown either flow separation, incipient separation or non-separated sheltering on the leeside portion for a range of wind-wave conditions, providing empirical justification for this assumption. Moreover, the assumption of displaced streamlines or separation on the leeward side causing negligible pressure force on the leeside is consistent with prior approaches used to developing analytical theories on air-wave interactions (e.g., \cite{Belcher_Hunt_1993,jeffreys1}. In order to account for this effect a Heaviside function $\text{H}(x)=  \frac{1}{2}(x -|x|)/x$ is introduced. For the argument of the Heaviside function, we use the projection between the relative velocity and the surface gradient $(\boldsymbol{u}-\boldsymbol{C})\cdot \boldsymbol{\nabla} \eta$. When the dot product is positive, the fluid velocity in a local frame moving with the surface points into the surface and raises the pressure.
Conversely, when it is negative, the relative fluid velocity is moving away from the surface and the pressure force on that part of the surface is set to zero.
Finally, then, the stress from the windward potential flow model (WPM) is given by:
\begin{equation}
\tau^{\rm wpm}_{i3}  = \frac{1}{\pi} \,|(\boldsymbol{u-C}) \cdot \boldsymbol{\hat{n}}|^2 \, |\boldsymbol{\nabla} \eta|^2 \,\, \text{H} \left[ (\boldsymbol{u}-\boldsymbol{C})\cdot \boldsymbol{\nabla} \eta \right] \, 
\hat{n}_i, \,\,\,\, i = 1,2.
\label{stress_heavi}
\end{equation}
Moreover, using the Heaviside function enables representation of the force from waves moving faster than the airflow. Then the pressure force-producing side of the wave would be on the negative slope side of the wave in contrast to the scenario of waves moving slower than the airflow.  

It is worthwhile to consider the situation when general phase velocity $\boldsymbol{C}$ may diverge ($|\boldsymbol{C}| \rightarrow \infty$) when  $|\boldsymbol \nabla \eta |^2 \rightarrow 0$ since  the surface gradient  is in the denominator in Eq. \ref{gen_vel}. Such a divergence in $C$ does not create problems for the model  since the pressure force is proportional to $\sim |\boldsymbol{\nabla} \eta|^{2}$, so that the resulting stress still vanishes even if $|C| \rightarrow \infty$ as $\sim |\boldsymbol{\nabla} \eta|^{-1}$. For airflow going over a locally flat surface (i.e. $\nabla \eta = 0$), there is no form drag.  
We emphasize that only the time and space dependence of the horizontally resolved height distribution  is required in this wall model. Moreover, note that the prefactor $1/\pi$ is determined from the theory with no further adjustable parameters. The model is expected to be applicable for small ramp angles $\alpha << \pi$ (say $\alpha <0.1 \pi$,  which translates to wave steepness of approximately $a < 0.4$). The other important element to consider is the location of the LES velocity $\textbf{u}$ used as input for the model. It should be chosen as a distance far enough away from a piecewise ramp of horizontal size $\Delta L$ so that the velocity can be approximated as being unaffected by that portion of the surface. We will chose a vertical distance equal to $\Delta L$ or equal to the horizontal LES grid spacing $\Delta x$.  

\subsection{Discussion}
A few comments contrasting the presently proposed windward potential flow model in Eq. \ref{stress_heavi} to the surface gradient drag (SGD) approach introduced in \cite{AndersonMeneveau2010}, and also used in \cite{aditya}, are pertinent. The SGD  model was based on the assumption that the entirety of the local incoming momentum flux translates into the form drag upon impinging on a flow obstacle, or roughness element \citep{AndersonMeneveau2010}. Hence, the direction of the force was given by the impinging momentum flux vector (or the relative velocity to the wave velocity in \cite{aditya}). Physically the assumption meant that the entirety of the incoming momentum would be transformed into a force. However, especially for small angles (or realistic wave steepness ranges), a significant fraction of the incoming momentum flux also leaves the control volume, thus decreasing the resultant drag force. Here, in the windward potential flow model we instead evaluate the pressure distribution on the surface. As a result, a different scaling is obtained (see below for more detailed discussion), and the direction of the force is determined only by the inclination of the surface and not by the relative momentum flux direction. 

Regarding the scaling of the stress due to  waves obtained from the 
windward potential flow model of  Eq. \ref{stress_heavi}, it is useful to estimate the order of magnitude of expected stress magnitude for a simple case: Assume a constant air velocity $u$ and phase-speed $c$ for a monochromatic wave moving in the $x$ direction with $\eta = a\cos{(kx - \omega t)}$. The average of $|\nabla \eta|^2$ (e.g. at $t=0$) over half wavelength (since only the forward-facing side generates drag as represented by the Heaviside function) is $\lambda^{-1}\int_{\lambda/2}^{\lambda} (ak)^2 \sin^2(kx) \, dx = (1/4)(ak)^2$ so that the mean stress is of the form $\overline{\tau}^{\rm wpm} \approx (4\pi)^{-1} (u-c)^2 \,(ak)^2$. This scaling is proportional to steepness square and a prefactor of order (0.1) is similar to the scaling of the pressure distribution obtained from wind-wave generation theory by \cite{jeffreys1}.

\subsection{Equilibrium Model For Fully Unresolved Surface Roughness}
\label{sec:eqwm}

As is usually done for wall modeled LES of high Reynolds number boundary layer flows, we can account for small scale interactions with unresolved surface features  at the air-sea interface with an effective roughness length $z_0$ based on the classic Equilibrium Wall Model  \citep{moeng1984}. A surface roughness length has been used to represent the drag effects of ocean surfaces on the ABL \citep{Deskos2022}. However, we also wish to represent the friction resulting from viscous action on a smooth surface in case there are no appreciable small-scale roughness elements. In order to efficiently describe smooth, rough and transitionally rough cases, it is convenient to use the formulation proposed in \cite{meneveau2020note}. There, the determination of the wall stress is formulated as a ``generalized Moody diagram fit'' appropriate for wall modeled LES. In this formulation, the wall stress is obtained using a wall modeling friction factor $c_f^{\rm wm}$ 

\begin{equation}
\tau^{\rm eqm}_{i3} = 
\frac{1}{2} \, c_f^{\rm wm}(Re_\Delta,z_0/\Delta)\, \,\,
%\left( \frac{Re_{\tau \Delta}^{\text{uf}}}{Re_{\Delta}}
%\right)^2\ 
\hat{U}_{\text{LES}}^\Delta\, \hat{u}_i^\Delta , \qquad i = 1,2,
 \label{eqwm_mod}
\end{equation}
where $\hat{u}_i$ is the input velocity from the LES at the wall model height relative to the wave surface and the subscripts $1, 2$ are again used to represent velocities in the streamwise and spanwise directions. For cases in which the effect of the wave orbital velocity \citep{Kundu} is taken into account, we can set $\hat{u}_i = u_i - u_{o,i}$, where $u_{o,i}$ is the orbital horizontal velocity and $u_i$ is the LES input air velocity.
The friction factor $c_f^{\rm wm}$ depends on $Re_\Delta = U_{\rm LES}^{\Delta} \Delta / \nu$,  a Reynolds number based on  $U_{\rm LES}^{\Delta}$, the local horizontal velocity magnitude at  height $\Delta$. The friction factor also depends on a dimensionless surface roughness ($z_o/\Delta$). For a smooth surface the latter vanishes, while for fully rough surfaces the dependence on $Re_\Delta$ disappears.  
The friction factor  can be expressed as follows (reorganizing expressions given in \cite{meneveau2020note}):
\begin{equation}
c_{f}^{\rm wm}(Re_{\Delta},z_o/\Delta) = 2\,   \left[\left(\frac{Re_{\tau \Delta}^{\rm s}}{Re_\Delta}\right)^6 + \left(\frac{1}{\kappa}\log\frac{\Delta}{z_0}\right)^{-6}\right]^{1/3},
\label{eq:cftotal}
\end{equation}
where $\kappa = 0.4$ is the von Karman constant and $Re_{\tau \Delta}^{\text{s}}=u_*^s \Delta/\nu$ is the wall stress friction Reynolds number for a smooth surface depending on the friction velocity $u_*^s$ that would exist for a given $U_{\rm LES}^\Delta$. 
For smooth surfaces the first term depending on $Re_{\tau \Delta}^{\text{s}}$ dominates. The smooth surface friction Reynolds number $Re_{\tau \Delta}^{\text{s}}$ has been fitted to results from numerical integration of the equilibrium model differential equation \citep{meneveau2020note}. The fit is given by
\begin{equation}
 Re_{\tau \Delta}^{\text{s}}(Re_\Delta) = 0.005^{\beta_1 - 1/2}Re_{\Delta}^{\beta_1}[1 + (0.005 Re_{\Delta})^{-\beta_2}]^{(\beta_1-1/2)/\beta_2},
 \label{eq:Retaufits}
\end{equation}
where  $\beta_1 = (1+0.155 Re_\Delta^{-0.03})^{-1}$ and $\beta_2= 1.7 - (1+36 Re_\Delta^{-0.75})^{-1}$. 

For fully rough conditions the second term in Eq. \ref{eq:cftotal} dominates and is used to represent the effects of unresolved roughness like small-scale sea surface features.  Unresolved sea surface roughness can be caused by several processes. Observable wrinkles on the water surface, commonly called ripples or parasitic capillary waves, are created in the initial stages of wind-wave generation \citep{lin2008} or on the windward faces of steep gravity waves \citep{perlin1993,longuet-higgins_1963}. Ripples represent a mechanism for extracting energy from the primary gravity wave through viscous energy dissipation at short scales and they act as a surface roughness creating momentum loss in the airflow. In the present study, the time evolution and interaction of parasitic capillary waves with gravity waves is not considered, however, their presence as drag facilitators is taken into account. Extensive experiments of turbulent flow over young wind-generated waves (ripples) done by  \cite{Geva}, show an existence of wall similarity in the spatially developing boundary layer. They have proposed a fully rough velocity profile for turbulent airflow above ripples, which relates the shape of the profile to the local characteristic wave height
\begin{equation}
u^+ = \frac{1}{\kappa} \ln{\biggl(\frac{z}{\eta^\prime_{\text{\rm rms}}}\biggr)} + 8.5,
\label{u+rip}
\end{equation}
where $\eta^\prime_{\text{\rm rms}}$ is the local root mean square of the ripples' height distribution.  
For an LES grid that cannot resolve such surface roughness features,  the corresponding roughness length scale
of Eq. \ref{u+rip} can be employed to model the subgrid-scale  roughness features as
\begin{equation}
z_{o} = \eta^\prime_{\text{\rm rms}} \, e^{-8.5 \kappa}.
\label{zo_rip}
\end{equation}
Using Eq. \ref{zo_rip} in the EQM expression (Eq. \ref{eqwm_mod}) enables representation of small scale ripple effects as a boundary condition in the wall modeled LES of airflow over waves.

\subsection{Summary of the Moving Surface Drag Model}
The wall stress due to the windward facing horizontally resolved wave features is superposed to the equilibrium model stress due with the unresolved surface features. Thus the total (MOSD) wall stress due to moving waves is given by 
$\tau^{\rm mosd}_{i3}=\tau^{\rm wpm}_{i3}+\tau^{\rm eqm}_{i3}$, or
\begin{equation}
\tau^{\rm mosd}_{i3} = 
  \frac{1}{\pi} |(\boldsymbol{u-C}) \cdot \boldsymbol{\hat{n}}|^2  |\boldsymbol{\nabla} \eta|^2 \, \hat{n}_i  \, \text{H} \Bigl[ (u_j-C_j)\frac{\partial \eta}{\partial x_j} \Bigr] \, + \, \frac{1}{2} \, c_f^{\rm wm}  \,\hat{u}_i^\Delta \, \hat{U}_{\text{LES}}^\Delta, \quad i = 1,2.
 \label{MOSD}
\end{equation}
In this equation, ${\bf u}$ is the LES velocity at a height $\Delta L$ (comparable to the LES horizontal resolution), ${\bf C}$ is the generalized surface horizontal speed  given by Eq. \ref{gen_vel}, $\eta(x,y,t)$ is the prescribed sea surface and $\hat{\bf n} = \boldsymbol{\nabla} \eta/|\boldsymbol{\nabla} \eta|$,  while $\hat{\bf u}$ is the LES velocity at the equilibrium wall model height $\Delta$ possibly with the wave orbital velocity subtracted, and $\hat{U}_{\text{LES}}^\Delta=|\hat{\bf u}|$ is the velocity's magnitude.

 It is also worth noting that the only input needed in our WPM component is the airflow velocity and the geometric information of the surface (i.e. the position and time-dependent elevation function $\eta(x,y,t)$). Therefore, the model's formulation is general enough to be applicable not only to sinusoidal wave shapes, but for any type of moving surfaces including multiscale surfaces like a random ocean spectrum wavefield. The explicit demonstration of the model's applicability for a surface spectrum is left for future work, however, in Sect. \ref{sec:generalapps} we present  cases with multiple wavelengths as a precursor to exploring such scenarios. Furthermore, it is important to reiterate the assumptions and limitations of the model. We introduced the MOSD model with explicit constraints, notably the restriction to waves with slopes that do not exceed 0.4, and the assumption of displaced streamlines on the leeward face of the wave causing negligible pressure forces there. In its current form, the model does not account for nonlinear wave-wave interactions and evolution or breaking wave conditions. It is crucial to emphasize that the model operates within the realm of one-way coupling, specifically predicting momentum transfer from waves to wind through form drag.

 Additionally, we clarify that the modeling of unresolved surfaces (ripples and viscous effects) through the EQM occurs locally throughout the entire surface and not only on the windward side of the waves. We envision that instantaneously the tangential viscous stress are non-negligible on either the windward or leeward sides. Support for this assumption comes from the experiments by \cite{buckley2020} and \cite{veron2007}, which show viscous tangential stress peaks near the wave crest and gradually decreases to a non-negligible value on the leeward side.

\section{Implementation of MOSD Model in LES of Wind over Waves}
\label{sec:les_num}
The MOSD model is implemented in a research LES code that uses spectral-finite differencing  to simulate ABL flows \citep{BouZeid}. The code  originated from the work of \cite{albertson}. The open-source version of the code (LESGO) is available on Github \citep{lesgo} and includes several subgrid stress parameterizations, wall models, and wind turbine representations using actuator disk or actuator line models. It can be run using fully periodic boundary conditions or the concurrent-precursor approach in \citep{stevens2014} as inflow generation technique. The code has been validated  by several previous studies \citep{BouZeid,stevens2014,ghanesh2022}. The governing equations and numerical method used are discussed in \ref{sec:gov_eq}, while the numerical setup for the various validation cases used to evaluate the MOSD model is described in \ref{sec:setup}.

\subsection{Governing Equations and Numerical Method}
\label{sec:gov_eq}
The LESGO code solves the filtered Navier-Stokes equations in rotational form to ensure mass and kinetic energy conservation. These equations are: 
\begin{equation}
\frac{\partial {{u}_i}}{\partial x_i} = 0,
\end{equation} 
\begin{equation}
\frac{\partial {{u}_i}}{\partial t} + {{u}_i} \biggl(\frac{\partial {{u}_i}}{\partial x_j} - \frac{\partial {{u}_j}}{\partial x_i}\biggr) = - \frac{1}{\rho}\frac{\partial P_{\infty}}{\partial x_1} - \frac{\partial {P}}{\partial x_i} - \frac{\partial \tau_{ij}}{\partial x_j} + \nu \frac{\partial ^2 u_i}{\partial x_j^2},
\label{momentum}
\end{equation} 
where ${u_i} = ({u}, {v},{w})$ are the filtered velocity components in the streamwise ($x$), lateral ($y$), and vertical ($z$) directions, respectively.
In Eq. \ref{momentum},  $\tau_{ij} = \sigma_{ij} -\frac{1}{3}\sigma_{kk} \delta_{ij}$ is the deviatoric part of the subgrid scale (SGS) stress tensor $\sigma_{ij}$. The quantity ${P} = {P}_{\ast}/\rho + \frac{1}{3}\sigma_{kk} + \frac{1}{2}{u_i}^2$ is the modified pressure, where the actual pressure ${P}_{\ast}$  divided by the ambient density $\rho$ is augmented with the trace of the SGS stress tensor and the kinematic pressure arising from writing the nonlinear terms in rotational form. The deviatoric component of the SGS stress tensor ($\tau_{ij}$) is modeled using the eddy viscosity concept:
\begin{equation}
\tau_{ij} = - 2 \nu_t {S}_{ij},
\end{equation}
where $\nu_t $ is the turbulent eddy viscosity and ${S}_{ij}$ is the resolved strain-rate tensor. The diffusivity variable $\nu_t $ is modeled as:
\begin{equation}
\nu_t = (C_S {\Delta})^2 \sqrt{{S}_{ij} {S}_{ij}},
\end{equation}
where $C_S$ is the Smagorinsky model coefficient and ${\Delta} = (\Delta x \Delta y \Delta z)^{1/3}$ with $\Delta x$, $\Delta y$, $\Delta z$ the respective $x,y,z$ grid spacings. For this study we use the Lagrangian dynamic scale-dependent model developed by \citep{BouZeid} to determine the coefficient $C_S$ dynamically.
Along the streamwise and spanwise directions the code uses a pseudo-spectral method, while a second-order centered finite difference method is used for discretization in the wall-normal direction. For time advancement, the second-order accurate Adams-Bashforth scheme is used. In order to prevent undesired artificially long flow structures, a shifted periodic boundary condition is used \citep{munters2016}. On the top boundary of the domain, a stress-free boundary condition is imposed. For the bottom boundary, the MOSD wall model described in Sect. \ref{sec:MOSD_model} is implemented.  Regarding the input velocity used in the equilibrium wall model, \cite{BouZeid} proposed that test filtering eliminates some undesired effects that arise from using the equilibrium model, which is valid for mean flow. Following \cite{BouZeid}, the input velocity used in the EQM is test-filtered at a scale $2\Delta$. For simplicity, the air velocity input used for the MOSD model is also test-filtered at the same scale. Moreover, to reduce the log-layer mismatch, \cite{KawaiLarsson} proposed that the wall model height should be taken further away from the wall. Therefore, for  the EQM input, velocities are taken at a height $z = 2.5\Delta z$, corresponding to the 3rd vertical grid point in the LESGO code.  

The ABL flow is forced only in the streamwise direction by an imposed pressure gradient:
\begin{equation}
- \frac{1}{\rho}\frac{\partial P_{\infty}}{\partial x_1} = \frac{u_*^2}{h_{\rm bl}},
\end{equation}
where $h_{\rm abl}$ is the height of the computational domain. Additionally, these two variables are used to non-dimensionalize the system variables. All simulations are run until the flow reaches a statistically steady state, as determined by various statistics including the change in kinetic energy over time.
\subsection{Simulation Setup}
\label{sec:setup}
\begin{figure}
  \centerline{\includegraphics[width=12cm]{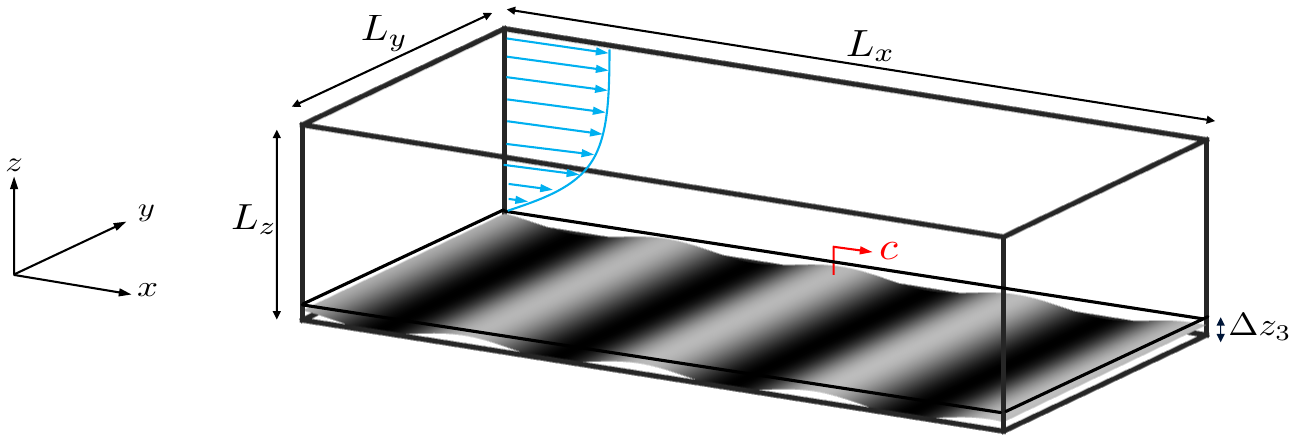}}
  \caption{Schematic representation of the computational domain used to simulate three-dimensional turbulent flow over monochromatic waves. The wave is moving in the positive streamwise direction with a phase velocity $c$. As described in Sect. \ref{sec:setup}, the entire wave lies below the grid point used for the wall model at a distance $\Delta$ from the $z=0$ plane (in our simulations we use the third grid-point, i.e., $\Delta = \Delta z_3$)}
\label{fig:wave_compdomain}
\end{figure}

The computational domain is shown in Fig. \ref{fig:wave_compdomain}. The prescribed wave motion shown is a monochromatic wave (later more general cases will be considered),    with a   wave amplitude $a$, wavelength $\lambda$, and wave phase speed $c$. Figure \ref{fig:wave_compdomain} also shows the wall model height, the height of the third vertical grid point $\Delta z_3$, where the air velocity is taken for the equilibrium portion of the model. For applicability of the EQM, the wave cannot be higher than $\Delta z_3$, therefore we enforce $\max{[\eta]}=a\leq 0.99\Delta z_3$, where $\max{[\eta]}$ is the amplitude $a$ for a monochromatic wave.
The vertical computational domain height $L_z$ is set equal to $h_{\rm bl}$ and all length-scales are normalized by $h_{\rm bl}$ such that $L_z=1$. The streamwise $L_x$ and spanwise $L_y$ domain sizes are chosen such that at least a minimum of 4 waves exist in the domain and $\Delta z/\Delta x = \Delta z/\Delta y$ is not smaller that $1/6$. Moreover, each wavelength is resolved in the horizontal direction by a minimum of 8 grid points. 

For validation of the MOSD model in LES, we have selected seven previous studies;  five are numerical and two are experimental.
These particular validation cases were selected because they cover a meaningful range of wave parameters and contain sufficient information to perform comparisons with our LES. Data from computational studies by  \cite{Wang_2021}, \cite{Zhang2019}, \cite{cao_shen_2021} and \cite{Hao2021} are used. They perform WRLES of turbulent airflow over a smooth monochromatic wave at the bottom boundary using a time-dependent boundary fitted grid. We also compare with numerical data from \cite{wu_popinet_deike_2022} who solve fully coupled wind forced water waves using DNS with a geometric volume of fluid method to reconstruct the interface. The experimental datasets used are from \cite{buckley2020} and \cite{yousefi_2020}. They performed high-resolution Particle Image Velocimetry (PIV) measurements of the turbulent airflow in close vicinity of the wavy interface.

\begin{table}
\centering
\caption{Wave and flow parameters, computational domain size and numerical resolution for all validation cases. Superscript "+" denotes normalization using the overall friction velocity $u_{*}$. Streamwise, spanwise and vertical domain sizes ($L_x, L_y, L_x$) are normalized by the boundary layer height $h_{\rm bl}$.} 
\label{tab:1}       
\begin{tabular}{cccccc}
\hline\noalign{\smallskip}
$ak$   & $c^+$ & $Re_\tau$ & $(L_x \times L_y \times L_x)/h_{\rm bl}$   & $N_x \times N_y \times N_z$   & Reference  \\
\noalign{\smallskip}\hline\noalign{\smallskip}
0.10   & 2.00 &1133    & $2\pi \times \pi  \times 1$   & $52  \times 26  \times 50$ & \cite{Wang_2021}\\
0.14  & 19.3 &5000 & $6 \times 3  \times 1$ & $52  \times 26  \times 50$ & \cite{Hao2021}\\
0.14  & 7.70 &5000  & $6 \times 3  \times 1$  & $52  \times 26 \times 50$ &    \%  \\
0.10   & 7.25 &1100 & $4\pi \times 2\pi  \times 1$  & $72  \times 36  \times 35$ &  \cite{Zhang2019}\\
0.10   & 23.8 &1110 & $4\pi \times 2\pi  \times 1$  & $72  \times 36  \times 35$ &    \% \\
0.06  & 6.57 &1320 & $6.35 \times  3.17\times 1$   & $52  \times 26 \times 50$ &  \cite{buckley2020}\\
0.12  & 3.91 &3025 & $5.93 \times  2.97 \times 1$ & $48  \times 24  \times 50$ &  \% \\
0.17  & 2.62 &5673 & $6.33 \times  3.16\times 1$ & $52  \times 26  \times 50$ &  \% \\
0.20  & 1.80 &9598 & $6.55 \times  3.27\times 1$  & $40  \times 20  \times 35$ &   \% \\
0.26  & 1.53 &10588 & $6.98 \times  3.49 \times 1$  & $40  \times 20  \times 35$ &  \cite{yousefi_2020}\\
0.20  & 8.00 &720  & $2\pi \times \pi  \times 1$   & $52 \times 26  \times 49$ & \cite{wu_popinet_deike_2022}\\
0.15  & 3.46 &434 & $6 \times  3 \times 1$  & $72  \times 36 \times 35$ & \cite{cao_shen_2021} \\
0.15  & 15.4 &384 & $6 \times  3 \times 1$  & $72  \times 36 \times 35$ &   \%  \\
0.15  & 54.3 &332 & $6 \times  3 \times 1$  & $72  \times 36 \times 35$ &   \%  \\
\noalign{\smallskip}\hline
\end{tabular}
\end{table}

The data from \cite{buckley2020} and \cite{yousefi_2020} were obtained at a large fetch such that the wavefield was sufficiently developed and reached fetch-limit equilibrium state. Furthermore, for all of their experimental cases, parasitic capillary waves are non-existing or their induced effect on the airflow is negligible. Therefore, for all validation cases, a smooth monochromatic wave is assumed (i.e. $z_{o} = 0$). Table \ref{tab:1} summarizes the important wave parameters, as well as the domain size and numerical resolution used in our WMLES for each validation case.

\section{Model Validation}
\label{sec:validation}

In order to test the new wall model, LES of turbulent boundary layer flow over monochromatic waves are conducted and directly compared against numerical and experimental data shown in Table \ref{tab:1}. 

\subsection{Turbulence Statistics: Mean Velocity and Reynolds Stress Profiles}
\label{sec:turb_stat}

 \begin{figure}
\centering
  \centerline{\includegraphics[width=12.4cm]{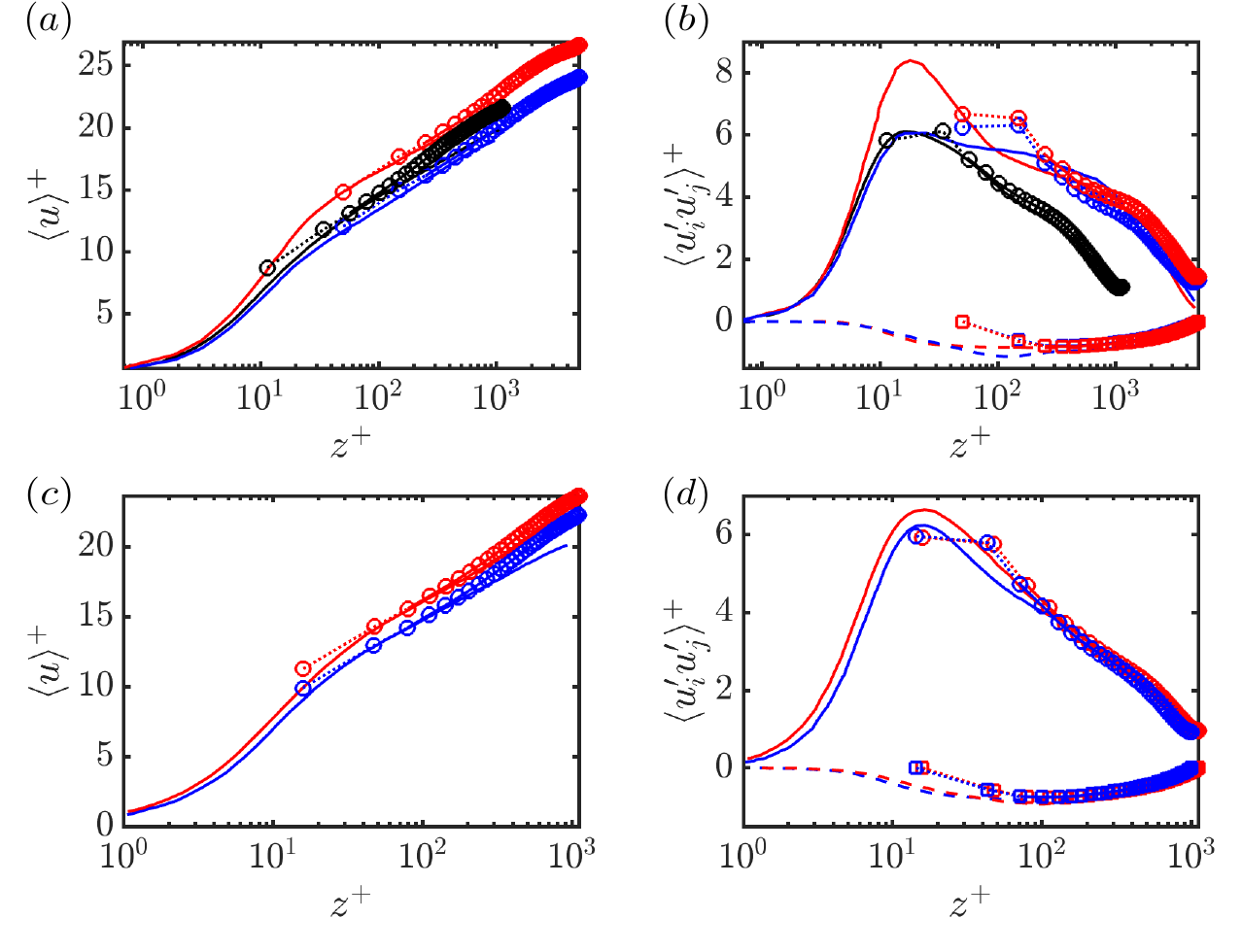}}
  \caption{(a,b) Comparison of mean streamwise velocity profiles and Reynolds stress profiles, respectively, from MOSD-based WMLES (symbols) and reference data using WRLES from \cite{Wang_2021} and \cite{Hao2021} (lines). 
  In both (a) and (b), Blue: \cite{Hao2021}, $ak=0.14\,| c^+=7.7$, Red: \cite{Hao2021}, $ak=0.14\,|\, c^+=19.3$ and Black: \cite{Wang_2021}, $ak=0.1\,| c^+=2.0$. (c,d) Comparison of mean streamwise velocity profiles and Reynolds stress profiles, respectively, from the MOSD wall modeled LES (symbols) and reference data using WRLES from \cite{Zhang2019} (lines). In both (c) and (d), Blue: \cite{Zhang2019} $ak=0.1\, |\,c^+=7.25$, Red: \cite{Zhang2019} $ak=0.1\,|\, c^+=23.8$. In (b) and (d) solid lines and circles are the streamwise Reynolds stress  $-\langle u^\prime u^\prime \rangle^+$ while dashed lines and squares are for the turbulent shear stress profiles $-\langle u^\prime w^\prime \rangle^+$. Values are normalized using the rough-wall friction velocity. For all figures, first 3 grid points are "inside" the wave, therefore they are shown jointed with a dotted line}
\label{fig:3}
\end{figure}

We compare first order and second order turbulence statistics of turbulent airflow above waves modeled using the MOSD approach in wall modeled LES.
For some cases, due to data availability the second order statistics are not compared.   First, in Fig. \ref{fig:3} we compare LES using the MOSD wall model with the WRLES data from \cite{Wang_2021}, \cite{Hao2021} and \cite{Zhang2019}.
In these LES  monochromatic waves are represented by a moving boundary-fitted grid on the bottom of the numerical WRLES domain. For all of these cases, the mean streamwise velocity profiles from the wall modeled LES using the MOSD approach agree well with the WRLES datasets. The velocity profiles follow the law of the wall and compare well with the data in the logarithmic region, with a small overshoot in the wake region near the upper portion of the profile (where the effects of the wall model are negligible and instead results depend mostly on the subgrid scale model \citep{wang2020comparative}, grid resolution and grid aspect ratio \citep{lozano2019error}). Several wave ages ($c^+$) are included in the comparisons of Fig.  \ref{fig:3}. A wave can be considered to be a high velocity wave (swell) if the wave age $c^+$ is larger than 25 \citep{Smedman1994,Deskos2021}. The case of \cite{Zhang2019} where $c^+ = 23.77$ and the \cite{Hao2021} case where $c^+ = 19.3$ can be considered   moderate wave age cases. For moderate wave ages the wave velocity can become comparable to the wind velocity, thus, drag is greatly diminished leading to larger mean velocity when scaled with friction velocity. This  trend is observed for the high $c^+$ case in the \cite{Hao2021} data in Fig. \ref{fig:3}c. For faster wave age conditions, the wave velocity is usually higher than the airflow velocity (such a case will be examined later), thus, the wave can energize the flow instead of acting to deplete momentum in the air.
In terms of second-order statistics (panels (b) and (d) in Fig.\ref{fig:3}),  there is  good agreement between the present results and the reference datasets except for the points nearest to the bottom surface, where some overestimation can be observed. 

\begin{figure}
  \centerline{\includegraphics[width=8.4cm]{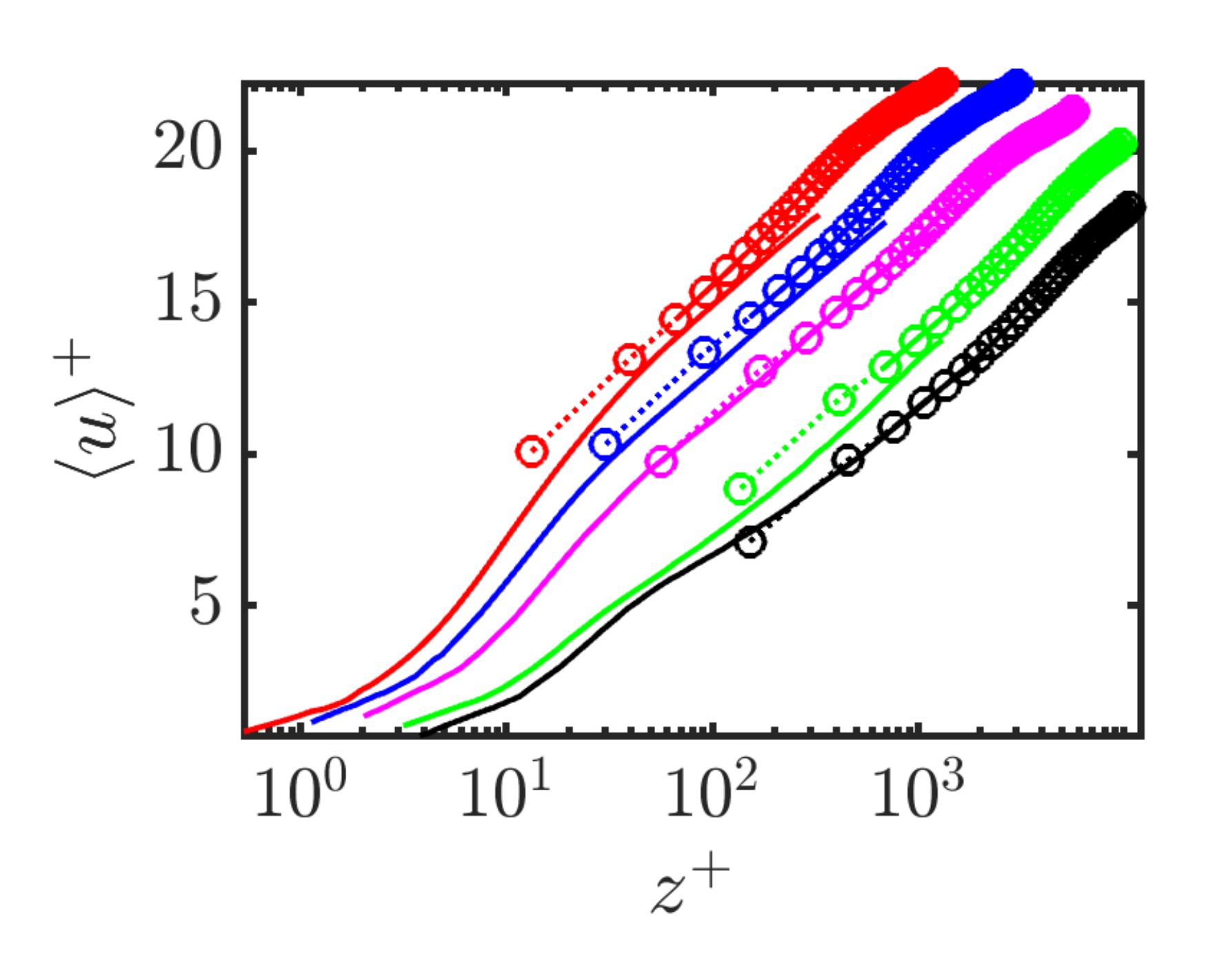}}
  \vspace{-5mm}
  \caption{Comparison of mean streamwise velocity profiles MOSD wall modeled LES (circles) and reference data from various experiments (lines) reported in \cite{buckley2020} and \cite{yousefi_2020}. Colors correspond to $ak = 0.06 \: |\: c^+ = 6.57$ (red), $ak = 0.12\: |\: c^+ = 3.91$ (blue), $ak = 0.17\: |\: c^+ = 2.62$ (magenta), $ak = 0.20\: |\: c^+ = 1.80$ (green), $ak = 0.26\: |\: c^+ = 1.53$ (black). First 3 grid points are "inside" the wave, therefore they are shown jointed with a dotted line}
\label{fig:4}
\end{figure}
Figure \ref{fig:4} shows a comparison of mean velocity profiles from MOSD-based WMLES and the various experimental datasets from \cite{buckley2020,yousefi_2020}. The LES correctly reproduces the  approximately logarithmic behavior and the amount of drag imparted  to the wind (i.e., it provides good predictions of the velocity reduction below smooth surface behavior) for all conditions considered. A slight overprediction of the velocity profile (slight underprediction of the drag experienced due to the waves) is visible for two of the cases ($ak = 0.12\: |\: c^+ = 3.91$ and $ak = 0.20\: |\: c^+ = 1.80$) but considering the range of conditions covered and the fact that no parameters are being adjusted from case to case, we consider the agreement quite good. 

Next, we compare our MOSD-based wall modeled LES with the DNS data generated by \cite{wu_popinet_deike_2022}, where wind wave growth was investigated using a two-fluid method.  In their work, the wave shape is initialized as a third-order Stokes wave whose parameters evolve as the wind and wave interact over time. We choose to validate against their $ak=0.2$ and $c^+=8$ case over their specific non-dimensional time range ($\omega t = 24 - 120$), where the wave parameters and form drag remain relatively constant.  For our LES, we specify a wave form $\eta$ according to:
\begin{equation}
\eta(x,y,t) = a\cos{(kx -\omega t)} + ak\,\eta_1(x,t) + (ak)^2\eta_2(x,t),
\end{equation}
where, $\eta_1(x,t),\eta_2(x,t)$ are first-and second-order correction terms (see \cite{Deike_Popinet_Melville_2015})

\begin{figure}
  \centerline{\includegraphics[width=12.4cm]{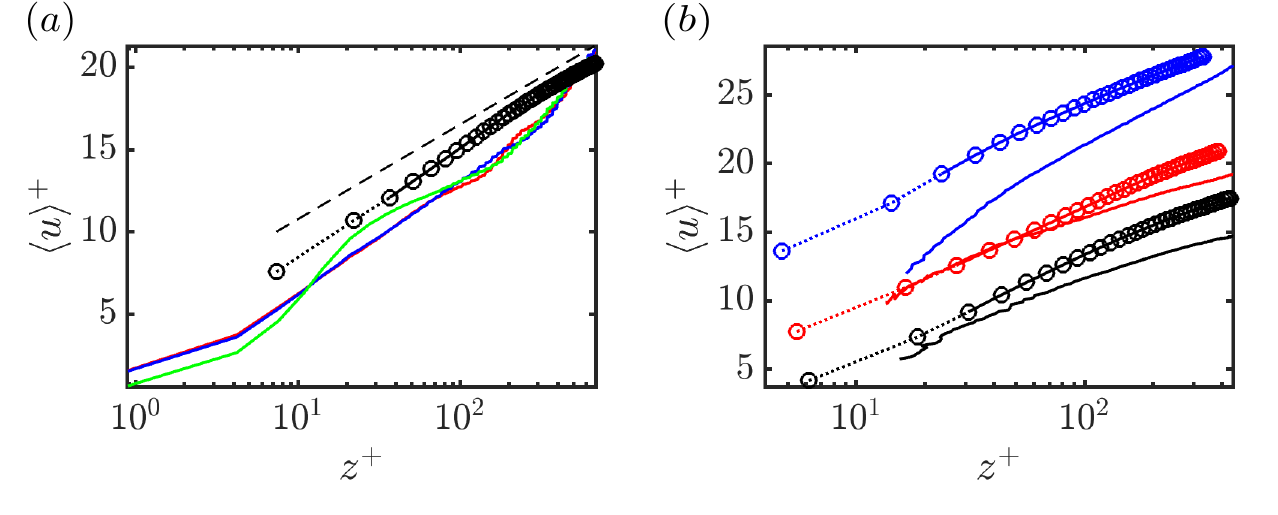}}
    \vspace{-4mm}
  \caption{(a)  Comparison of mean streamwise velocity profiles from MOSD-based WMLES (circles) and reference data (lines) using DNS from \cite{wu_popinet_deike_2022}. Line colours correspond to $\omega t = 24$ (red), $\omega t = 56$ (blue) and $\omega t = 120$ (green). The dashed line corresponds to the standard smooth surface case $\kappa^{-1}\log(z^+)+B$ with $\kappa=0.4$ and $B=5$. (b) Comparison of mean streamwise velocity profiles from MOSD-based WMLES (circles) and reference data (lines) using WRLES from \cite{cao_shen_2021}. Colors correspond to $ak = 0.15 \: |\: c^+ = 3.46$ (black), $ak = 0.15\: |\: c^+ = 15.38$ (red) and $ak = 0.15\: |\: c^+ = 54.30$ (blue).  For all figures, first 3 grid points are ``inside'' the wave, therefore they are shown jointed with a dotted line}
\label{fig:5}
\end{figure}

Figure \ref{fig:5}a shows the comparison of the mean streamwise velocity profile from our LES with the \cite{wu_popinet_deike_2022} DNS data. In this case the agreement is relatively poor, and it appears that the WMLES underpredicts the mean drag force acting on the wind.  Considering that the study done by \cite{wu_popinet_deike_2022} fully resolves all scales and coupling effects occurring at the air-water interface and that the their velocity profiles are not streamwise-averaged, one may expect some differences between the LES and the DNS data. Furthermore, we include a smooth profile (dashed line) to the Fig.~\ref{fig:4}b just as reference showing that the WMLES based on MOSD produces additional drag from the wave but insufficiently compared to this particular DNS. 

\begin{figure}
  \centerline{\includegraphics[width=12.4cm]{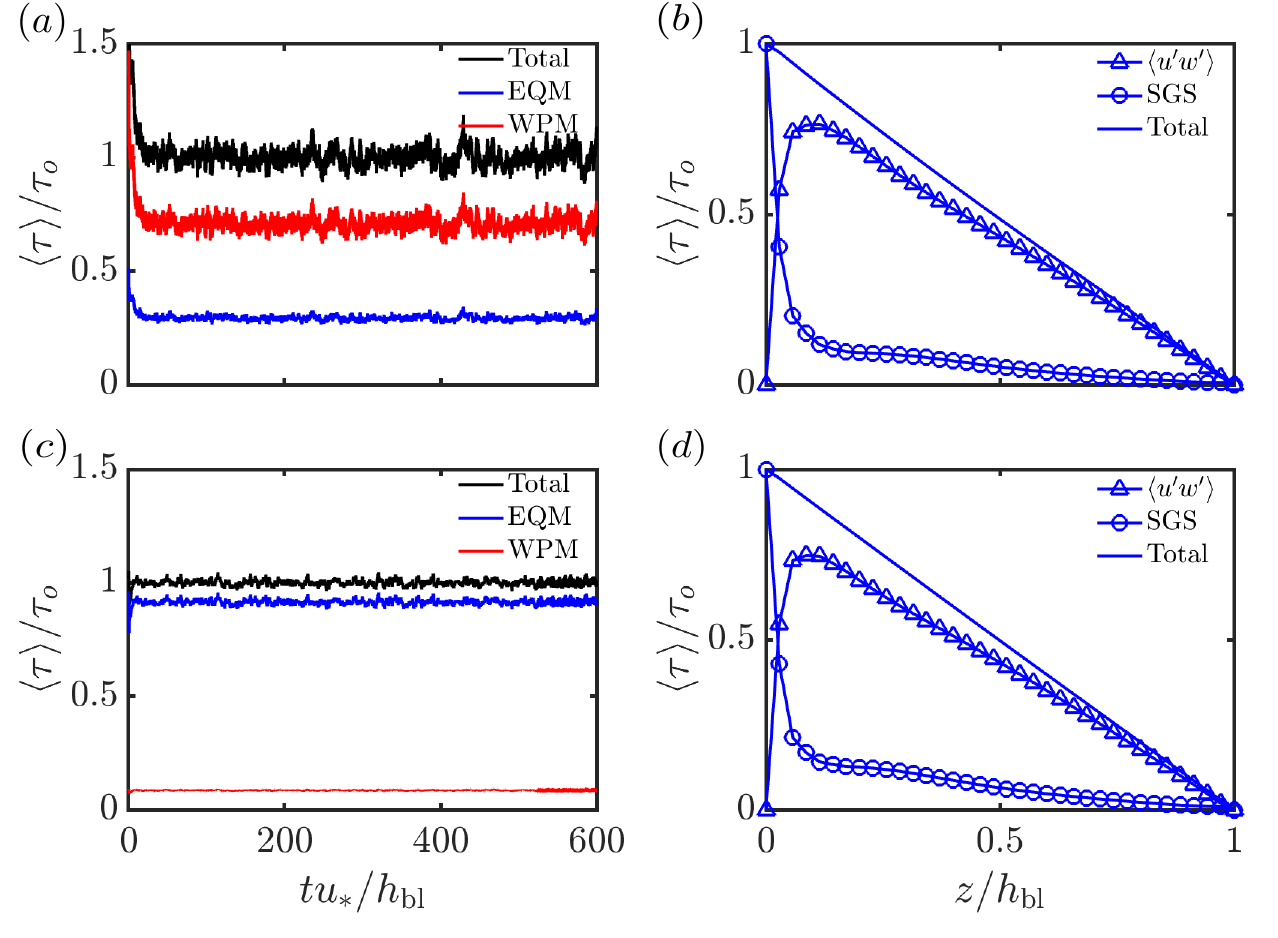}}
    \vspace{-4mm}
  \caption{(a,c) Time signals of the normalized MOSD wall stress components (Eq.\ref{MOSD}). The components are averaged in the $x,y$ plane. (b,d) Mean vertical shear stress budget from the MOSD-based WMLES. The (a,b) panels are for the reference case of \cite{yousefi_2020} and (c,d) panels are for \cite{Zhang2019} $ak=0.1\,| \, c^+=7.25$ case}
\label{fig:6}
\end{figure}

In order to test the performance of the MOSD wall model for the case of fast moving waves that impart momentum to the wind above, we compare with data from \cite{cao_shen_2021}. They use WRLES with a boundary fitted grid to solve for turbulent airflow over propagating waves with various phase speeds in a Couette flow setup. They include three cases of  a slow, moderate and fast moving wave. The three cases have the same wave steepness $ak=0.15$  to isolate the effects of the wave age. We simulate only the lower half of their domain (i.e. we consider $h_{\rm bl}$ to be the half-channel scale). Results are shown in Fig.~\ref{fig:5}b.  For the intermediate case, the model correctly predicts the velocity profile in the inner part of the profile, i.e., a reduction in drag as the wave age approaches the moderate regime. Differences near the channel center are to be expected since we are not simulating Couette flow. Furthermore, for the fast moving wave (c+ = 54.3) the MOSD-based WMLES predicts the addition of momentum to the wind. However, for this extreme condition the model overpredicts the momentum input from the wave to the wind, which leads to an over prediction of the velocity in the log-law region compared to the data. To the best of the our knowledge there is no experimental or numerical evidence of separation or sheltering on the windward side for fast moving wave conditions, therefore the assumption of displaced streamlines or airflow separation in the MOSD model may be invalid and possibly lead to an overprediction of negative form drag (e.g., a thrust). Future work should be focused on further improving the MOSD model to better predict the momentum transfer from fast-moving waves.  The slow wave age case also presents some differences between our WMLES and the comparison DNS data. It is worth noting that the data from \cite{cao_shen_2021} is at a moderate Reynolds number ($Re_\tau = 330 - 430$) for which additional effects may need to be considered. Still, one can argue that the MOSD model  predicts the main qualitative trends quite successfully.  

To illustrate the behavior of the MOSD model components for different wave parameters as well as their level of temporal fluctuations, Fig. \ref{fig:6} shows time signals of the total wall stress alongside the EQM and WPM components, normalized by the mean total wall stress ($\tau_o = u_*^2$). Only two cases are shown, which represent widely different cases in terms of wave parameters (see Table \ref{tab:1}). For both these cases the  total wall stress fluctuates around its mean  unit value, verifying that the balance between the wall stress and the nondimensional imposed pressure gradient. In Fig.~\ref{fig:6}a, the WPM dominates and represents approximately 60$\%$ of the total wall stress. This is expected because of the high wave steepness of this case ($ak=0.26$), corresponding to \cite{yousefi_2020}. Conversely, for the \cite{Zhang2019} case (Fig.~\ref{fig:6}c) the contribution of the WPM component is approximately 1$\%$ of the total stress. In this case, although the wave velocity is higher compared to the case presented in \cite{yousefi_2020}, for the specific Reynolds number of the airflow and with a wave steepness that is 2.6 times smaller, the contribution of the WPM would effectively be very small. The Reynolds and subgrid scale shear stress from MOSD-based WMLES for both reference cases are shown in Fig.~\ref{fig:6}b and Fig.~\ref{fig:6}d. For both, the dominant contribution is the resolved stress starting from 0 at wall and increasing linearly towards the half-channel height. The subfilter stress decreases from the wall towards the half-channel height, and starts becoming negligible in terms of the total stress around $z/h_{\rm bl} \approx 0.6$.

\begin{figure}
  \centerline{\includegraphics[width=12.2cm]{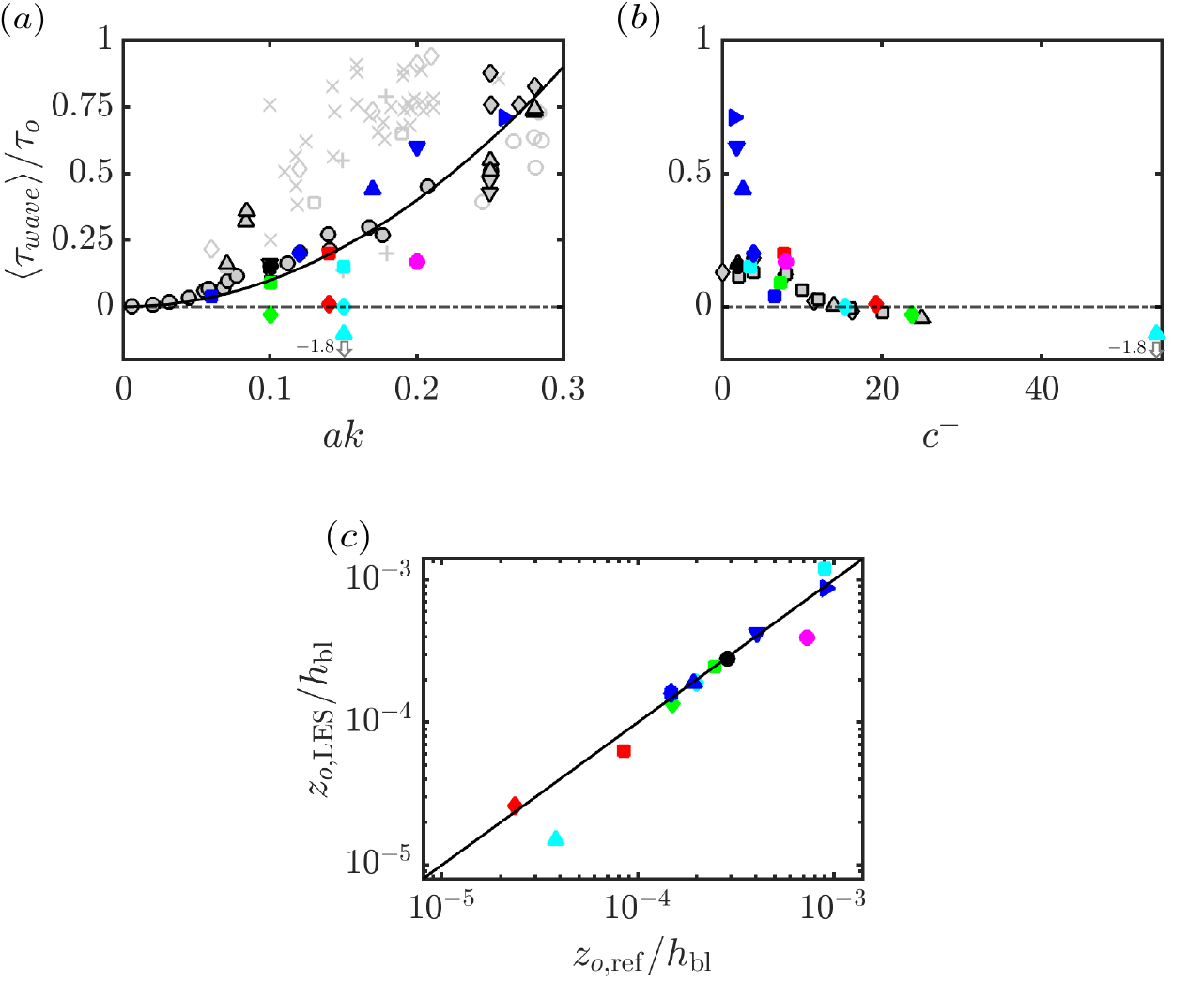}}
  \caption{ (a) Fraction of wave form drag as a function of wave steepness: comparison of MOSD-based wall modeled LES and data from the literature. Solid crosses $(\protect\greycross)$ : \cite{Grare_2013}; Solid circles $(\!\!\protect\blackbordergrayfilledcircle)$  : \cite{peirson_garcia_2008}; Diamonds $(\protect\greyedgediamond)$ : \cite{buckley2020}; Squares $(\protect\greyedgedsquare)$ : \cite{Funke2021}; Plus signs $(\protect\greyplus)$ : \cite{mastenbroek_1996}; Solid diamonds $(\protect\filledgreyedgediamond)$ : \cite{banner_1990}; Empty circles $(\protect\greyedgedcircle)$ : \cite{banner_peirson_1998}; Solid stars $(\protect\filledgreypentagram)$ : \cite{kihara2007}; Downward-pointing triangles $(\protect\greyfilledblackedgetriangledownward)$ : \cite{yang_shen_2010} and Upward-pointing triangles $(\protect\greyfilledblackedgetriangle)$  : \cite{Sullivan_2018}. The black line corresponds to the $\sim (ak)^2$ scaling. Full colored symbols correspond to the MOSD-based WMLES for the following comparison datasets: Black circle $(\!\!\protect\blackfilledcircle)$ : \cite{Wang_2021}. Red square $(\protect\redsquarefilled)$ : \cite{Hao2021} $ak=0.14\,|\, c^+=7.7$. Red diamond $(\protect\reddiamondfilled)$ : \cite{Hao2021} $ak=0.14\,|\, c^+=19.3$. Green square $(\protect\greensquarefilled)$ : \cite{Zhang2019} $ak=0.1\,|\,c^+=7.25$. Green diamond $(\protect\greendiamondfilled)$ : \cite{Zhang2019} $ak=0.1\,|\, c^+=23.8$. Magenta circle $(\!\!\protect\magentafilledcircle)$ : \cite{wu_popinet_deike_2022}. Blue square $(\protect\bluesquarefilled)$ : \cite{buckley2020} $ak = 0.06 \,|\, c^+ = 6.57$. Blue diamond $(\protect\bluediamondfilled)$ : \cite{buckley2020} $ak = 0.12\,|\, c^+ = 3.91$. Blue upward-pointing triangle $(\protect\bluetrianglefilled)$ : \cite{buckley2020}  $ak = 0.17\,|\, c^+ = 2.62$. Blue downward-pointing triangle $(\protect\bluetriangledownfilled)$ : \cite{buckley2020}   $ak = 0.20\, |\, c^+ = 1.80$. Blue right-pointing triangle $(\protect\bluetrianglerightfilled)$ : \cite{yousefi_2020} $ak = 0.26\, |\, c^+ = 1.53$. (b) Fraction of wave form drag as a function of wave age: comparison of MOSD-based wall modeled LES and data from the literature. Solid diamonds $(\protect\filledgreyedgediamond)$ : \cite{sullivan_2000}; Solid squares $(\protect\greyfilledblackedgesquare)$ : \cite{kihara2007} and Solid triangles $(\protect\greyfilledblackedgetriangle)$ : \cite{yang_shen_2010}. (c) Comparison between the surface roughness estimated from MOSD-based wall modeled LES and the surface roughness from the validation cases. For the WMLES, $z_o$ is obtained from fitting the log law region with a least square method. Filled color symbols are as in (a)}
\label{fig:7}
\end{figure}
 
In order to place our WMLES results using MOSD in the context of  results available in the literature, we show the stress fraction from the windward potential flow model normalized by the total wind stress plotted as function of wave steepness $(ak)$ alongside data from a range of prior studies in  Fig. \ref{fig:7}a. The stress fraction has often been argued to scale as $(ak)^2$ for small wave steepness and the proposed windward potential flow model follows this scaling since for monochromatic waves $(ak)^2 \sim \langle (\nabla \eta)^2 \rangle $. However, the literature shows significant scatter since the stress fraction not only depends on wave steepness but also on wave age, Reynolds number and other effects. As is visible in Fig. \ref{fig:7}a  our MOSD-based wall modeled LES similarly displays significant scatter and does not fall on a single $(ak)^2$ line. 
Figure \ref{fig:7}b shows the normalized wave form drag as a function of the wave age also compared to cases in the literature. We observe how the wave drag fraction decreases as the wave age increases, for both the MOSD-based wall modeled LES as well as for the available literature data. Figure \ref{fig:7}c shows a  comparison of the surface roughness from the WMLES and the reference datasets. The solid line is where $z_{o,{\rm LES}} = z_{o,{\rm ref}} $. The $z_{o,{\rm LES}}$ was obtained by fitting the data in the log-law region ($0.03 h_{\rm bl} - 0.1h_{\rm bl}$) using a least square method with prescribed slope ($1/\kappa$ and we use $\kappa=0.4$) to determine the intercept. We observe a good agreement between the MOSD-based WMLES and the majority of the comparison data. 

\subsection{Phase-averaged and Wave-induced Motions}
\label{sec:wave_indu}

The prior validation analysis based on mean velocity and Reynolds stress profiles did not specifically highlight the phase-resolving features of the model. For that purpose, we next employ  phase-averaging \citep{hussain_reynolds_1970} to compare results from the MOSD-based WMLES to data from prior phase resolving studies. Phase averaging allows us to isolate and analyze the specific motions and structures that are induced by the presence of surface waves. The triple decomposition of an arbitrary random signal \citep{hussain_reynolds_1970} is given by:
\begin{equation}
q(x,y,z,t) = \langle  q \rangle (z) + \tilde{q}(x,z,\phi) + q^\prime (x,y,z,t),
\end{equation}
where $\langle  q \rangle$ is the ensemble average component, $\tilde{q}$ denotes the wave-induced component that depends on phase $\phi$, and $q^\prime$ is the uncorrelated (random) turbulent component. The average $\langle  q \rangle$ is obtained by averaging over all horizontal directions and time, i.e., over $(x,y,t)$. 
The  wave-induced component   $\tilde{q}$ is given by $\tilde{q} = q_\phi  - \langle  q \rangle $, where $q_\phi$ is computed from: 
\begin{equation}
q_\phi (x,z,\phi) = \frac{1}{N} \sum_{n=0}^{N} \langle q(x,y,z,(\phi/2\pi+n) \, T ) \rangle_y,
\end{equation}
where $T$ is the period of the wave, and the subscript $y$ denotes averaging over the spanwise direction.   

\begin{figure}
  \centerline{\includegraphics[width=12.5cm]{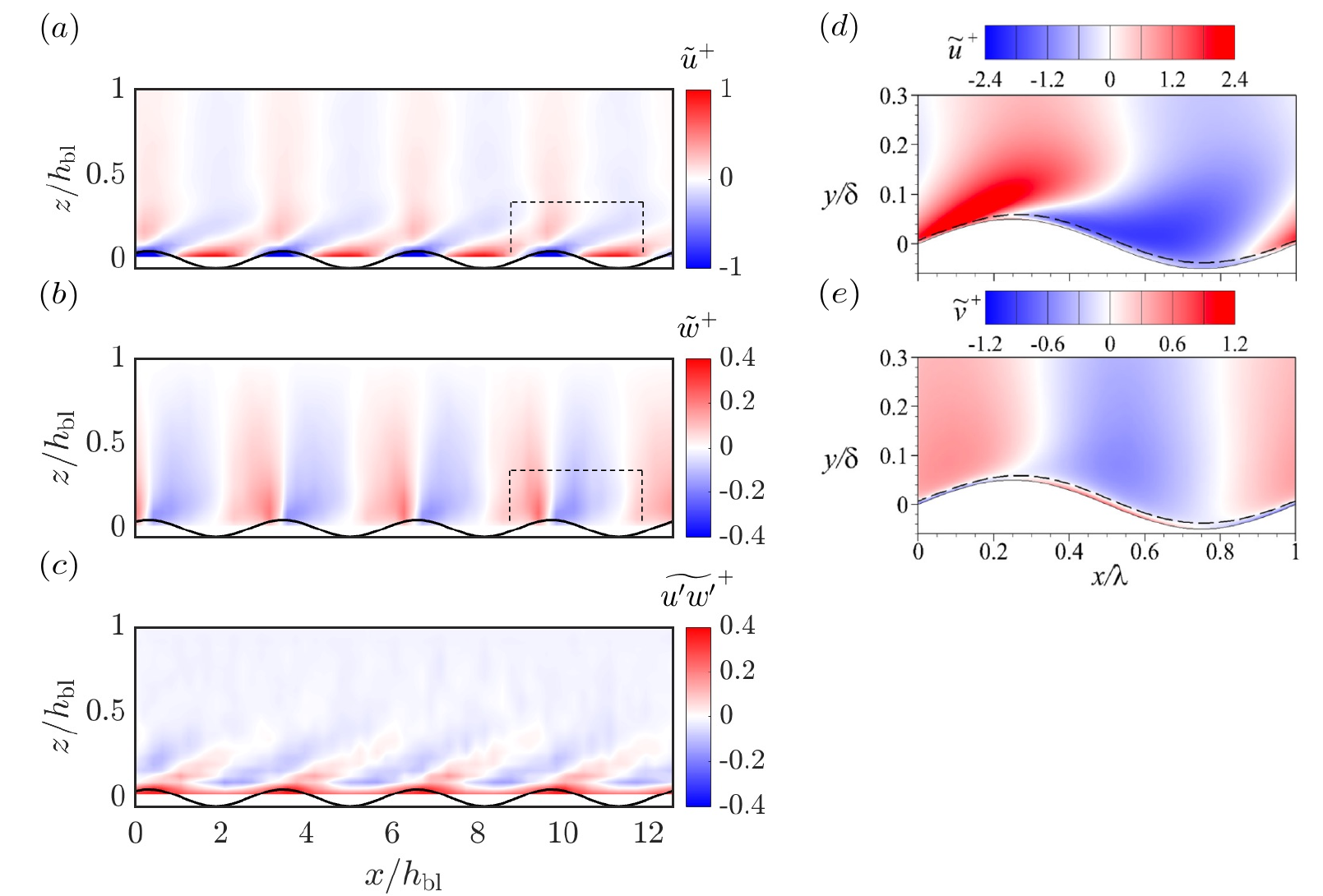}}
  \caption{Normalized wave-induced motions for \cite{Zhang2019} validation case. (a,d) Horizontal velocity from MOSD-based WMLES and WRLES comparison data from \cite{Zhang2019}
  (reproduced with permission), in a small region near the wave shown using dashed lines in (a), respectively. (b,e) Vertical velocity from WMLES and WRLES, respectively. (c) Phase averaged Reynolds shear stress from WMLES. Dashed line is the same viewing size as WRLES data}
\label{fig:8}
\end{figure}
Contours of the normalized wave-induced horizontal and vertical velocities calculated by the MOSD-based WMLES for the \cite{Zhang2019} validation case are shown in Fig. \ref{fig:8}a and Fig.  \ref{fig:8}b, respectively. Results are shown for the phase corresponding to the wave position shown in the figure. Here, the horizontal and vertical induced motions are positive on the windward side and negative on the leeward side of the wave. This behavior indicates that the airflow above accelerates as it passes over the windward face and later decelerates as it goes towards the leeward face, which is clearly consistent with the so-called ``pumping''effect \citep{sullivan2008large}. The flow also accelerates directly above the crest \citep{buckley2016}. 

These results indicate that the wave-induced structures of the horizontal and vertical velocities simulated in MOSD-based WMLES agree qualitatively with the results presented in \cite{Zhang2019}. However, it can be seen that the WMLES predict an earlier onset of acceleration very near the surface in the troughs/valleys compared to the  \cite{Zhang2019} WRLES comparison data, which instead show  a negative streamwise wave-induced velocity near the wave. Also, we note that the amplitudes of the phase-averaged motions are underpredicted compared to the data by a factor of approximately 2-3. The MOSD model is built to be used specifically as a boundary condition for the horizontal velocity component, hence the vertical velocity is not directly affected by the boundary condition. The observed vertical motions in Fig.\ref{fig:8}b are a result of the divergence-free condition of the flow. When the airflow approaches the windward side of the wave, volume conservation  dictates that the flow must move upwards. While this effect is not reproduced at a quantitative level, it is reproduced qualitatively and compared to the full velocity, the discrepancy is relatively small (on the order of $u_*$ only). 

Finally, we examine the turbulent stress caused by the presence of the waves (wave-induced turbulent stress). The wave-induced turbulent stress is defined as:
\begin{equation}
\widetilde{u^\prime w^\prime} =  (u^\prime w^\prime)_\phi - \langle u^\prime w^\prime \rangle.
\end{equation}
where, $(u^\prime w^\prime)_\phi$ is the phase-averaged component and $\langle u^\prime w^\prime \rangle$ is the horizontal and time average of the shear stress. The wave-induced Reynolds shear stress obtained from MOSD-based WMLES for the \cite{Zhang2019} case is shown in Fig. \ref{fig:8}c. A positive stress appears very near the surface of the wave on the windward side, extending towards the crest. On the leeward side of the wave, negative shear stress is observed, creating a positive-negative pattern along the wave crests. The negative values are tilted and are found always downwind of the wave, the positive values are also tilted and can extend towards the windward side of the next wave. This positive-negative pattern extends vertically until $z/h_{\rm bl} \approx 0.5$. The wave-induced turbulent shear stress behavior found using the MOSD-based WMLES shows good qualitative agreement with the trends obtained in the experiments by \cite{yousefi_2020}.  At the wavy surface, the horizontal momentum exchange is given by viscous stress and wave-induced stress. The latter has been shown to be exactly the form drag at the surface \citep{phillips_1957}. Therefore, to provide a more in-depth understanding of the momentum transfer caused by the MOSD model, we investigated the wave-induced momentum flux (\(\tilde{u_1}\tilde{w_1}\)) (figures not shown). The analysis shows that the peak of wave-induced momentum transfer occurs on the windward side of the wave, near the crest, with a smaller secondary peak immediately above and slightly downstream of the wave crest. The peak concentration of momentum flux is consistent with the fundamental assumption of the WPM model, which posits that the form drag is confined to the windward side of the wave.

The model validation results illustrate the performance of the MOSD wall modeling approach in LES: it can represent phase-averaged overall drag effects reasonably accurately over a range of wave parameters, and captures qualitative features as function of wave phase. This performance is achieved at significant computational savings compared to phase and wall-resolved LES. For example, in the WRLES \cite{Zhang2019} case, the streamwise grid size remains uniformly resolved with a resolution of $\Delta x^+ \approx 45$, while in the vertical direction, the grid size ranges from $\Delta z^+ \approx 0.05$ to $10$. In comparison,  LES with MOSD wall model for this case employs a grid size resolution of $\Delta x^+ \approx 190$ and $\Delta z^+ \approx 32$. This difference in resolution requirements translates to a significant reduction in computational cost, a factor of between $200-2{,}000$ (obtained as $(190/45)^3\times(32/10)-(190/45)^3\times(32/1)$, assuming a similar ratio in spanwise resolutions and also taking into account the reduced time-stepping requirements of WMLES). 

\section{Applicability Beyond Monochromatic Aligned Waves}
\label{sec:generalapps}
\begin{figure}
  \centerline{\includegraphics[width=10cm]{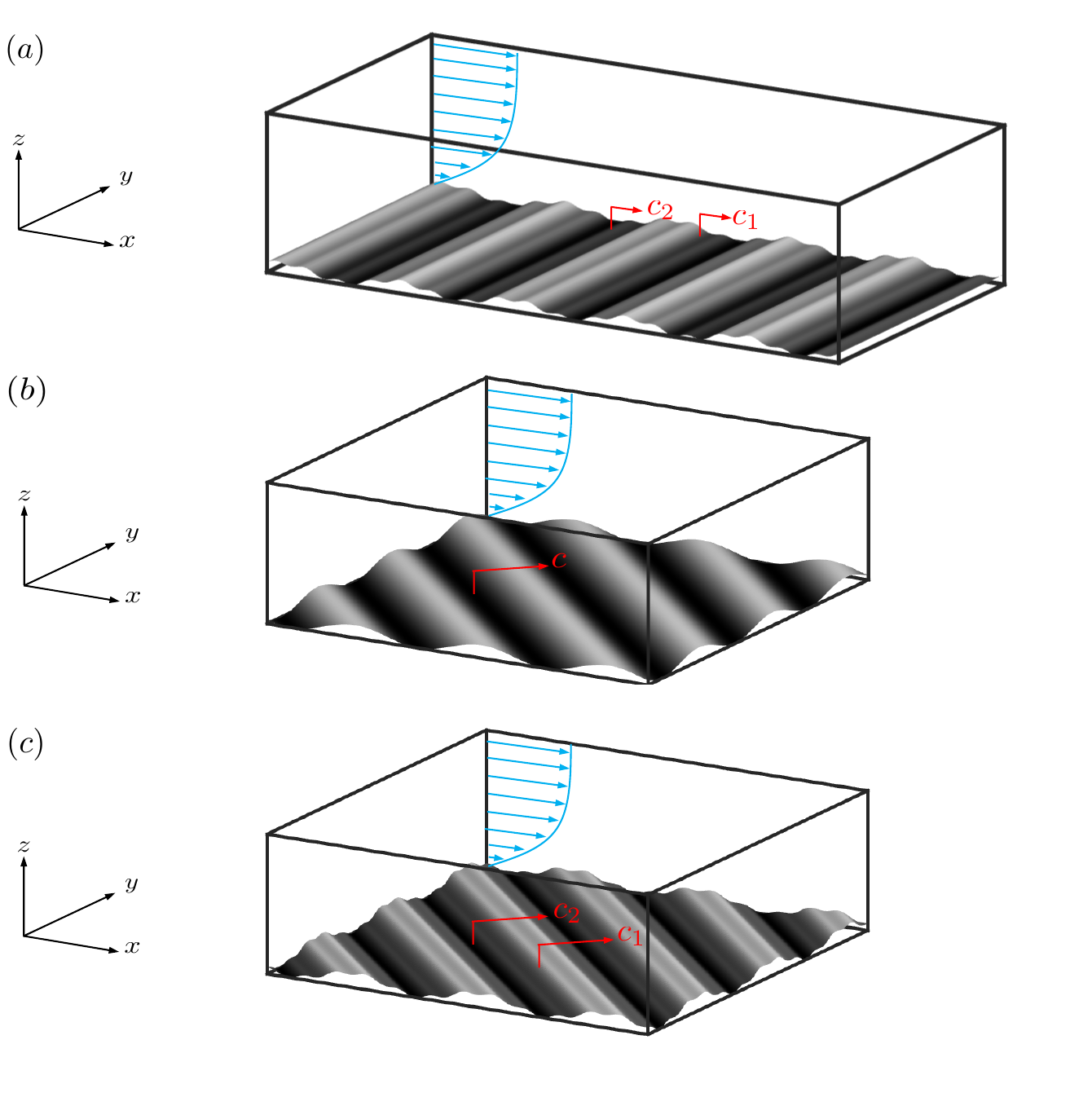}}
  \caption{Schematic representation of three wave scenarios used to illustrate the more general applicability of the MOSD wall model. (a) Two waves with different wavelengths moving at distinct velocities in the same direction as the wind. (b) One wave moving at an angle with respect to wind direction. (c) Two waves moving at distinct velocities at an angle with respect to the wind direction}
\label{fig:9}
\end{figure}

In the previous section we have shown how the MOSD-wall model can capture mean and phase effects for simple waves (monochromatic, single phase velocity aligned with the mean wind direction). In reality, ocean waves are multiscale, i.e., a superposition of many waves with different wave lengths and phase velocities. To capture the impact of such multiscale waves on the ABL, the MOSD model relies on the local slope as determined from $\nabla \eta$ and based on (\ref{gen_vel}) for evaluating the general velocity of the surface. It is therefore applicable in principle to any wave spectrum and wave phase velocity direction. To illustrate applicability of the MOSD model to waves beyond single wave-length, wind-aligned cases,  we choose three cases. The first scenario corresponds to two distinct waves with different wave steepness moving in the same direction but at different phase velocities. The second scenario is of a single wave moving at an angle relative to the wind (oblique waves) at a fixed phase velocity. The third scenario represents two distinct waves, both moving at an angle with respect to the wind and at different phase velocities. Figure \ref{fig:9} illustrates the three different scenarios. 

Table \ref{tab:2} summarizes the wave parameters, numerical domain and resolution for each of the cases considered. We choose to use the \cite{Zhang2019} case of $ak=0.1$ and $c^+ = 7.25$ as a base case for comparison purposes. For these illustrative applications of the MOSD wall model, there are no literature data for us to compare with, so the simulation results are shown to demonstrate that the model can be applied to such cases and that it predicts trends that are as expected and physically meaningful. 

\begin{figure}
  \centerline{\includegraphics[width=12.4cm]{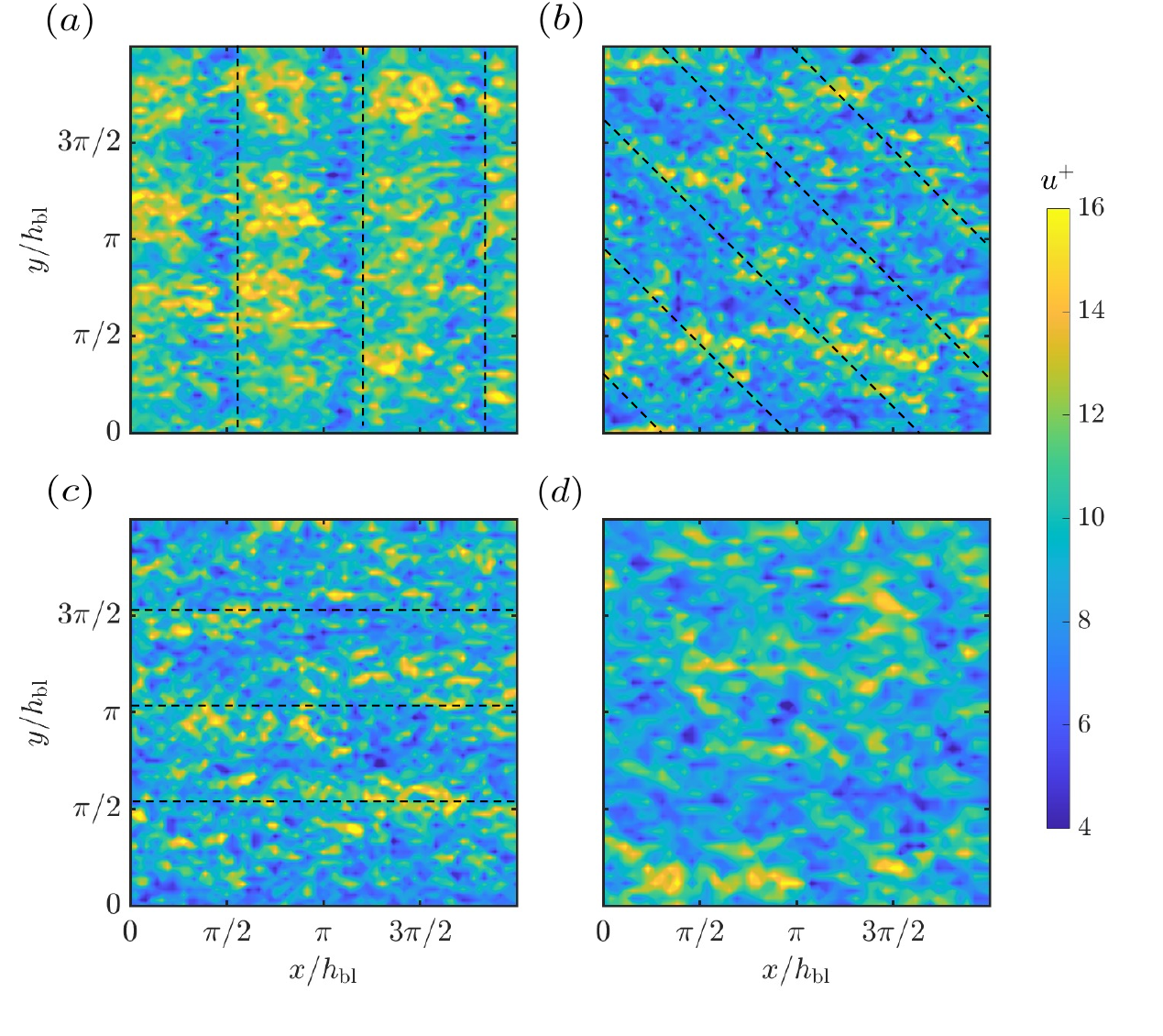}}
  \caption{Instantaneous streamwise velocity contours in the horizontal plane at $z= 0.0143 h_{\rm bl}$. (a) MOSD-based WMLES of \cite{Zhang2019} ($\gamma=0^\circ$). (b)  SMW45. (c) SMW90  and  (d) DMW. For reference, the wind is moving left to right. The dashed lines represent the crest of the waves}
\label{fig:10}
\end{figure}

For the first scenario of double aligned waves (DAW) we consider two cases.  For the first case (DAW07), the second wave has a small steepness of $ak=0.07$. For the second case (DAW15), the second wave has a bigger steepness $ak=0.15$. For both cases, the phase velocity of the second wave is the same. For the second scenario  we consider two cases of single misaligned waves (SMW). For the first case (SMW45) the wave moves at $45^\circ$ with respect to the mean wind direction. In the second case (SMW90), the wave moves at $90^\circ$, i.e., completely perpendicular to the mean wind direction. For the third scenario, where we have double misaligned waves (DMW), only one case is simulated. In this case, the second wave has the same parameters as DAW15 and both waves are moving at $45^\circ$ angle with respect to the incoming wind.  In all the cases shown in Table \ref{tab:2} have the same primary wave (wave 1) in terms of steepness and phase velocity (same as the \cite{Zhang2019} reference case). 
\begin{table}
\centering
\caption{Wave parameters, computational domain size and numerical resolution for MOSD-based wall modeled LES cases considered in \S \ref{sec:generalapps}. Subscript ``w1" and ``w2" correspond to wave 1 and wave 2, respectively. $\gamma$ represents the angle between the direction of the wave and the mean wind direction (x-direction).} 
\label{tab:2}       
\begin{tabular}{cccccccc}
\hline\noalign{\smallskip}
Case & $(ak)_{\text{w}1}$ & $(c^+)_{\text{w}1}$  &$(ak)_{\text{w}2}$   & $(c^+)_{\text{w}2}$ & $\gamma$ & $L_x \times L_y \times L_x$  & $N_x \times N_y \times N_x$ \\
\noalign{\smallskip}\hline\noalign{\smallskip}
    DAW07 & 0.1 &7.25 &0.07  & 1 & $0^\circ$ & $2\pi \times \pi  \times 1$  & $68  \times 34  \times 64$     \\
    DAW15 &  0.1 &7.25& 0.15  & 1 & $0^\circ$ & $2\pi \times \pi  \times 1$  & $68  \times 34  \times 64$   \\
    SMW45 & 0.1 & 7.25&0 & 0 & $45^\circ$ & $2\pi \times 2\pi  \times 1$  & $72  \times 72  \times 70$   \\
    SMW90 & 0.1& 7.25& 0 & 0 & $90^\circ$& $2\pi \times 2\pi  \times 1$ & $72  \times 72  \times 70$     \\
    DMW & 0.1 & 7.25 & 0.15  & 1 & $45^\circ$ &  $2\pi \times 2\pi  \times 1$ & $72  \times 72  \times 70$   \\
\noalign{\smallskip}\hline
\end{tabular}
\end{table}
To first visualize the effects of different wave scenarios onto the turbulence near the surface,  instantaneous contours of streamwise velocity component normalized by the friciton velocity at a wall distance height  of $z= 0.0143 h_{\rm bl}$ are shown in Fig. \ref{fig:10}. 
The presence of the waves is visible as a decrease in velocity in the form of bands, which in the near-surface region is correlated with the crests of the wave (as discussed in Sect. \ref{sec:wave_indu}, the trend is reversed and the flow preferentially accelerates above the crest of the waves). The flow deceleration is clearly visible for the case of wave aligned with the wind (Fig. \ref{fig:10}a) and slightly visible for the SMW45 case. For the SMW90 case (Fig. \ref{fig:10}c) there are no clear bands of deceleration which is expected since the only velocity component impinging onto the waves is the spanwise component ($v$) which is very small compared to the horizontal velocity ($u$). For the DMW case, the presence of waves is difficult to observe because this case has two waves of distinct sizes, however, some remnant effects of the primary wave can be observed. These visualizations also further demonstrate  the phase-dependent effects captured using the MOSD model. 

\begin{figure}
  \centerline{\includegraphics[width=12.4cm]{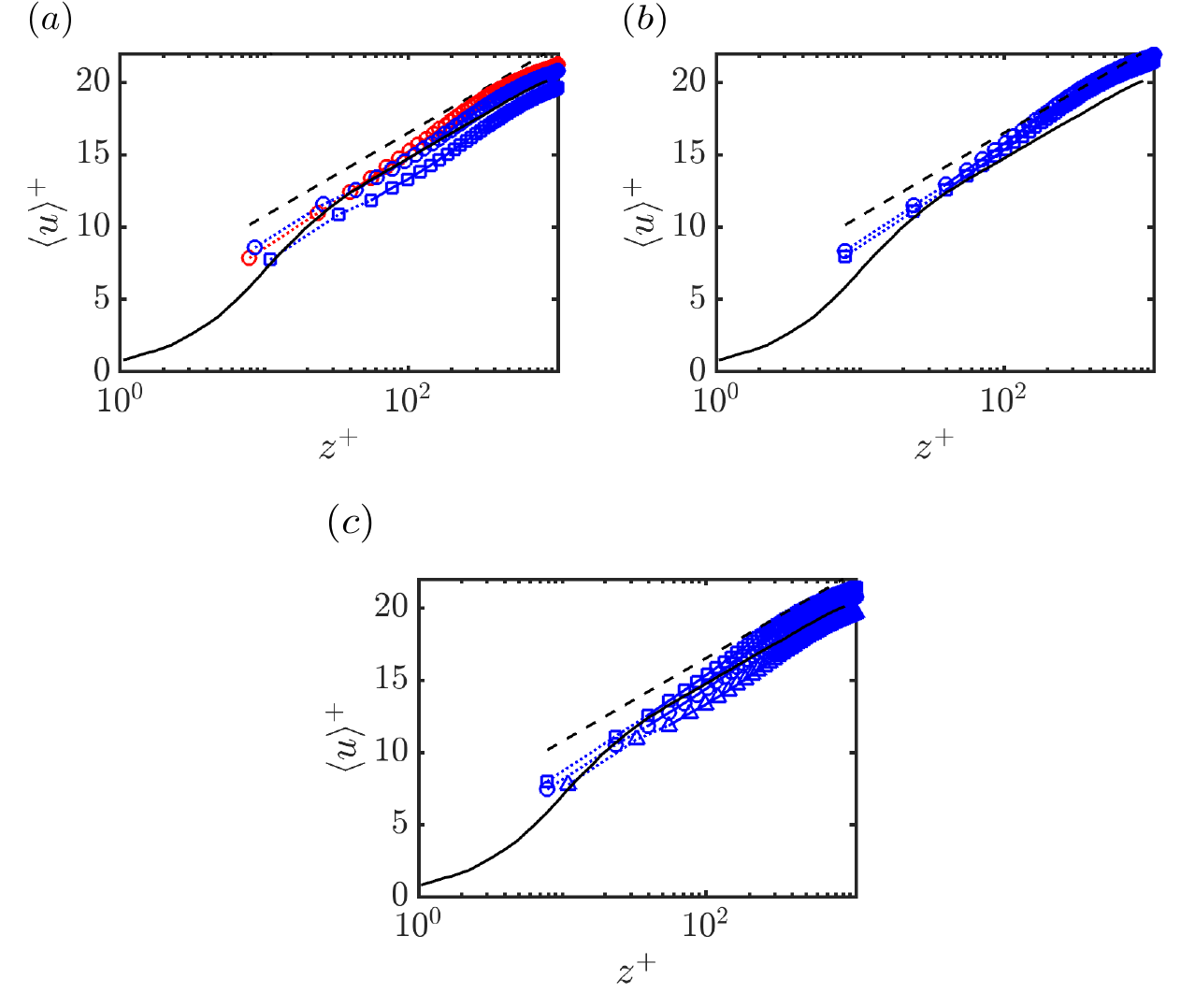}}
  \caption{(a) Mean streamwise velocity profiles for superposed two wave cases. Symbols are from MOSD-based WMLES. Blue circle: DAW07; Blue square: DAW15; Red circle: WMLES of \cite{Zhang2019} case;  (b) Mean streamwise velocity profiles for single misaligned wave cases. Symbols are from MOSD-based WMLES. Circle: SMW90 and Square: SMW45; (c) Mean streamwise velocity profiles for double misaligned waves cases. Symbols are from MOSD-based WMLES. Circle: DMW; Square: SMW45 and Triangle: DAW15; In all figures: the dashed line corresponds to the standard smooth surface case $\kappa^{-1}\log(z^+)+B$ with $\kappa=0.4$ and $B=5$. Solid line: WRLES from \cite{Zhang2019} for aligned single-mode wave shown as comparative case}
\label{fig:11}
\end{figure}
Figure \ref{fig:11}a shows the mean streamwise velocity profile for the first scenario consisting of two aligned waves (DAW07 and DAW15 cases). We expect that two waves would increase the amount of drag being imparted onto the airflow above. However, only the DAW15 case shows a clear offset of the velocity profile characteristic of an increase in drag.
For the DAW07 case, the drag imparted is virtually the same as the drag created from only one wave. This can be explained by noting that it's steepness is $30\%$ less than the primary wave, and since the drag scales by $ak^2$, the second wave becomes negligible in terms of drag. 

Figure \ref{fig:11}b shows the mean streamwise velocity profile comparison for cases SMW45 and SMW90. Here, we observe that there is less drag being imparted onto the airflow versus the wind and wave aligned case \cite{Zhang2019}, therefore a positive  offset in the log-layer region is observed. This decrease in drag is clearly due to the angle between wind and wave since both cases have the same wave parameters as the \cite{Zhang2019} case. This behavior is consistent with studies by \cite{karniadakis,tomiyama,Ghebali}, where transverse motions created by passive means like wall oscillation and transverse traveling walls have been found to create some drag reduction of turbulent boundary layer flows. 

The mean streamwise velocity for the DMW case is shown in Fig. \ref{fig:11}c. The drag generated differs slightly from that in the DAW15 case. Considering the results obtained in Fig. \ref{fig:11}b, it seems logical to infer that although the present case and DAW15 case have the same wave parameters, the slight difference in velocity profile offset is due entirely to the wave direction with respect to the wind. 

\section{Comparison with Wind Tunnel Data for 
Offshore Wind Turbine}
\label{sec:vwind_turbine_sim}
In the previous sections, we have  tested the accuracy and general applicability of  the MOSD wall model in  LES to  predict air flow over moving waves of various types. This wall-modeling  approach offers a cost-effective method for capturing the influence of waves on the ABL including phase-dependent information. In this section, we explore the applicability of the MOSD wall model to LES of offshore wind turbine wake flow. Specifically, a comparison with data from a recent laboratory experiments for the case of a fixed bottom  offshore wind turbine is presented. 

\subsection{Experimental Setup}
The experiments for a scaled fixed bottom wind turbine \citep{fercak_2022} were performed in the closed loop wind and water tunnel at Portland State University (PSU). The wind tunnel test section had a height of 0.8 m, width of 1.2 m and test-length of 5 m. The wind tunnel speed (free-stream velocity) was fixed at 6 m/s. The tunnel ceiling was configured to produce a nominally zero-pressure gradient boundary layer. The water tank covered the full wind tunnel floor and provided a water depth of 0.3 m. A wave paddle was positioned at the entrance of the test-section and was controlled by a stepper motor to produce scaled deepwater waves. Since the focus is solely on the effect of waves on wake development, an idealized setup using unidirectional waves was used to  focus the study on the flow interactions between turbulence and a single wavelength wave. Based on the wind tunnel size, a diameter of 0.15 m was selected for the scaled wind turbine, resulting in a geometric scaling ratio of 1:600 in comparison to a full scale turbine with (e.g.) a diameter of 90 m. The size of the turbine model and resulting Reynolds number is sufficiently high to show realistic wake dynamics \citep{chamorro2012}. The  water tank depth of 0.3 m corresponds to 180 m in full-scale. It is noted that a 180 m water depth is in a range relevant to floating wind turbines since it is well beyond the approximately 50 m limit for current fixed bottom turbines. A schematic representation of the experimental set up is included in Fig. \ref{fig:concurrentwave_compdomain}.

The model turbine thrust coefficient $C_T \approx 0.65$ was measured using a uni-axial load cell on the turbine rotor at the  tip speed ratio ($TSR = 5$) used in the experiment. The wave paddle was set to a frequency resulting in a wave with a measured period of ($0.8 \pm 0.01$) s. The measured wave speed ($c$) generated by the paddle was ($1.2 \pm 0.08$) m/s, and the measured wavelengths ($\lambda$) was ($1 \pm 0.06$) m. The water surface elevation (the interface between wind and wave) was measured with Light-Induced Fluorescence (LIF). Particle Image Velocimetry (PIV) was used to measure 2D velocity fields in streamwise aligned ($x-z$) planes. The friction velocity fitted from the mean velocity profile and using a von Karman constant of $\kappa =0.4$ was found to be $u_*=0.26$ m/s. The boundary layer height in the region of the turbine was approximately $h_{\rm bl} = 0.2$m. More detailed information regarding the experimental set up can be found in \cite{fercak_2022}. 

\subsection{MOSD-based WMLES Simulation Setup}
\begin{figure}
  \centerline{\includegraphics[width=12cm]{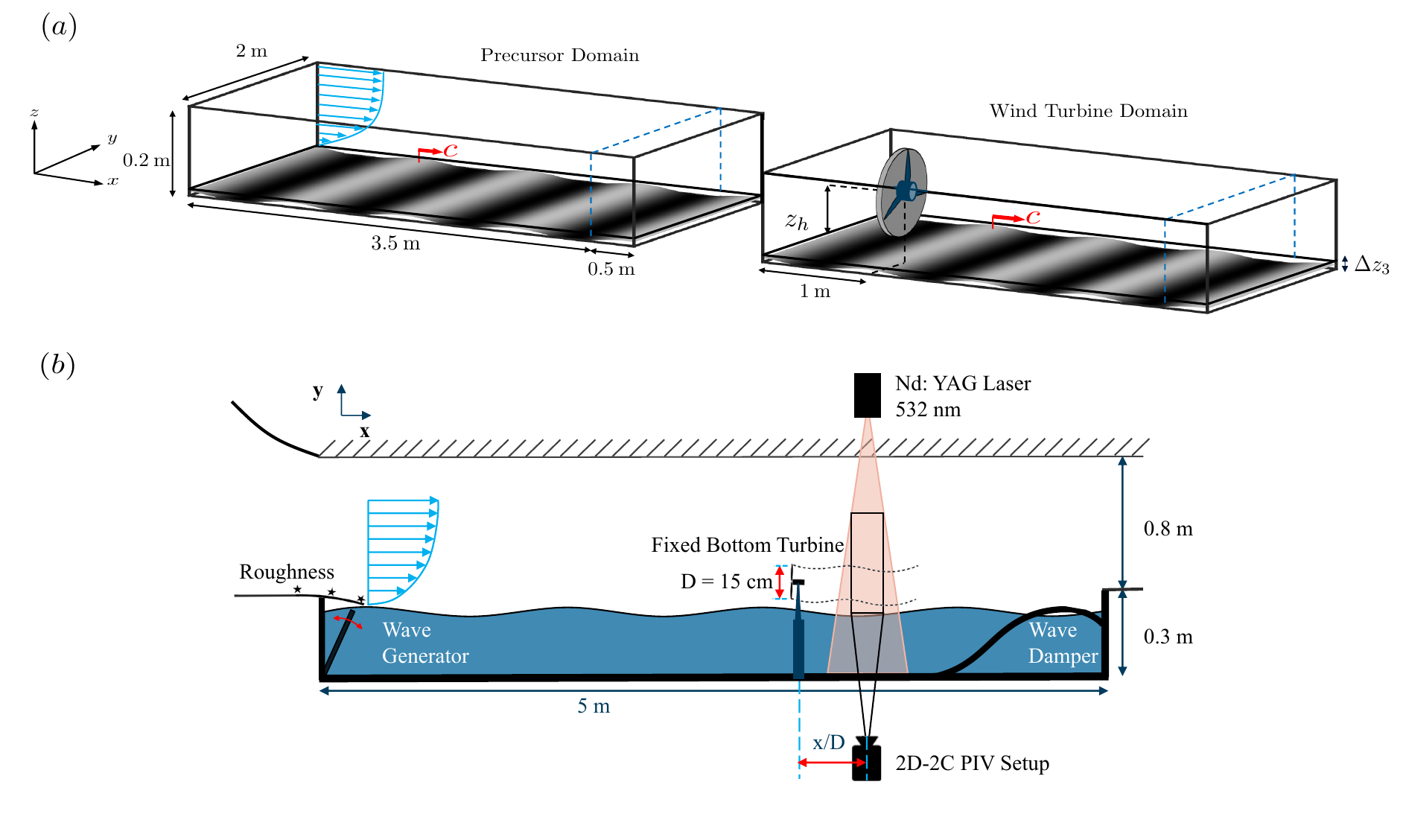}}
  \caption{(a) Schematic representation of the two computational domains (Precursor and Wind Turbine domain) used to simulate a fixed-bottom offshore wind turbine model in a wind tunnel experiment. In LES, the turbine and the nacelle are modeled using an actuator disk model at a hub height of $z_h= 0.115$ m. At each time step the data at the end of the first domain (section starting from the blue dash line), is copied to the end of the second domain with the wind turbine where it acts as a turbulent inflow. In the wind turbine domain, the fringe region starts at the blue dash line and ends at the end of the computational domain. (b) Schematic representation of the experimental set up}
\label{fig:concurrentwave_compdomain}
\end{figure}

We simulate wind-wave-wake dynamics numerically using the 
same procedure described in \ref{sec:setup} for wind-wave interaction with a wind turbine introduced into the numerical domain. Figure \ref{fig:concurrentwave_compdomain}a shows the computational domain used.The effect of the wind turbine rotor on the wind is modeled using the actuator-disk model (ADM), which is commonly applied to the study of land-based wind farms \citep{calaf2010large,shapiro2019} and offshore wind farms \citep{yang2014large,yang_offshorefarm_2014}. We use a local thrust coefficient (allowing us to use the disc-averaged air velocity instead of a far upstream reference velocity \citep{meyers2010AIAA}) of $C_T' = 1.15$, which is computed from $C_T'=C_T/(1-a)^2$, where $a=1/4$ is used to represent the turbine rotor induction \citep{calaf2010large}. We also include a model for the drag caused by the model turbine nacelle at the grid points near the turbine center using a drag coefficient of $C_D = 12$ (this value was selected aiming to create a velocity deficit near the turbine similar to that of the experiment). The simulation of wind-wave-turbine is performed using two computational domains, a precursor and a wind turbine domain \citep{stevens2014}. The turbine is placed in the wind turbine domain while the turbulent inflow conditions are generated in the precursor domain. The domain height (half-channel scale in the idealized version of the experiment) is set to equal the boundary layer height, i.e. $L_z=h_{\rm bl}=0.2$m. The usual air viscosity $\nu = 1.5 \times 10^{-5}$ m$^2$/s is used and the Reynolds number of the wind tunnel experiments is matched by selecting the forcing pressure gradient to correspond to the measured friction velocity $u_*=0.26$ m/s. 

The MOSD wall model is applied to both the precursor and wind turbine domains. A shifted periodic boundary condition \citep{munters2016} is used in the streamwise direction of the precursor domain to prevent artificially long flow structures from developing. In the concurrent-precursor method \citep{stevens2014} the turbulence and velocity developed in the precursor is then used as an inflow condition to the wind turbine domain. At at each time step, a part of flow from the precursor domain is copied to the outflow region of the wind turbine domain. A fringe region is then defined to smoothly transition between the wind turbine flow and the region of flow copied from the precursor domain and the prescribed ``outflow'' in the wind turbine domain is then equal to the inflow to this domain owing to the periodic boundary conditions used in the pseudo-spectral method in the horizontal directions.

\begin{table}
\centering
\caption{Wave parameters, computational domain size and numerical resolution for MOSD-based wall modeled LES of the flow of the   \cite{fercak_2022} experiment} 
\label{tab:3}       
\begin{tabular}{cccccccc}
\hline\noalign{\smallskip}
   $ak$   & $c^+$ & $(L_x \times L_y \times L_x)/h_{\rm bl}$   & $N_x \times N_y \times N_z$ & $h_{\rm bl}$ (m) & $\frac{z_{o}}{h_{\rm bl}}$ \\
\noalign{\smallskip}\hline\noalign{\smallskip}
    0.08   & 4.44     & $20 \times 10  \times 1$  & $116 \times 108  \times 36$&   0.2 & $1.29\times 10^{-4}$\\
\noalign{\smallskip}\hline
\end{tabular}
\end{table}

Table \ref{tab:3} summarizes the wave parameters from the experiment and the numerical resolution used for the wall-modeled LES. The experiments exhibited small wind driven waves (ripples) on top of the primary wave which was generated by the paddle. These small scale waves could not be horizontally resolved with the current LES grid, therefore, they are modeled as a surface roughness using Eq.\ref{zo_rip}. By means of light-induced fluorescence (LIF), the r.m.s size of the ripples was measured to be $\eta'_{\text{r.m.s}} = 2.58\times 10^{-5}$ m , which represents roughly $\approx$ 6\% of the amplitude of the wave-maker generated wave. Since we want to include these unresolved ripples as a surface roughness in addition to including viscous effects on the surface, we use Eq. \ref{eqwm_mod} with $c_f^{\rm wm}$ determined using Eqs. \ref{eq:cftotal} and \ref{eq:Retaufits}. 

\subsection{Results}
\begin{figure}
  \centerline{\includegraphics[width=12.4cm]{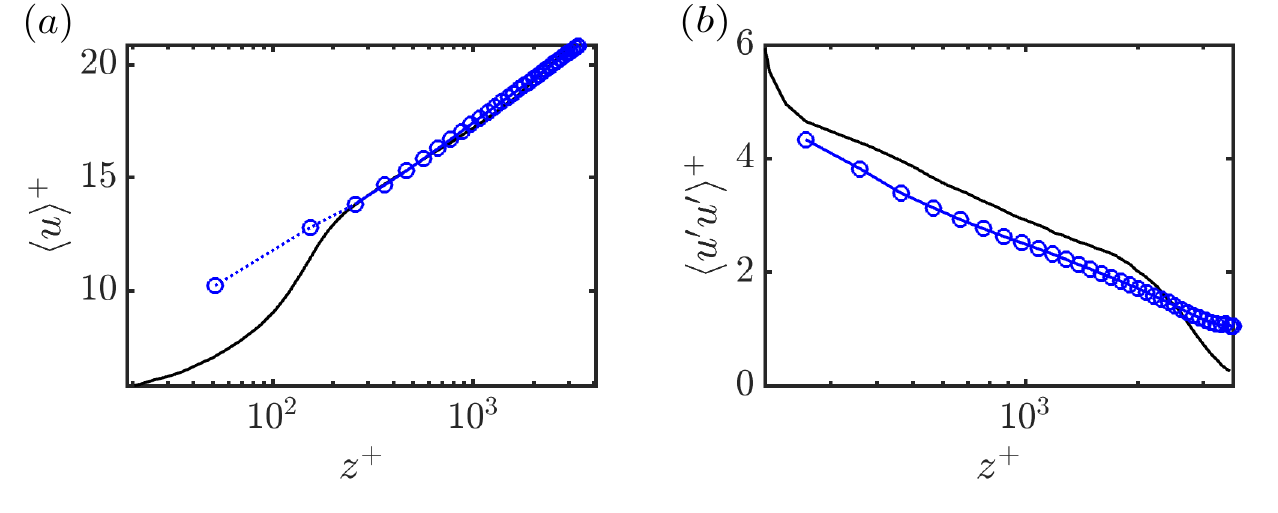}}
  \caption{Comparison of MOSD-based WMLES of air flow over water waves in a wind tunnel experiment (reference without wind turbine). (a) Mean streamwise velocity profiles. (b) Mean streamwise Reynolds stress profile. For both figures, circles show results for LES  and lines show the experimental data from \cite{fercak_2022} }
\label{fig:13}
\end{figure}
First, a LES of only wind over waves using the MOSD wall model was undertaken to verify that  the turbulent inflow conditions from LES match those of the experiments. Figure \ref{fig:13} shows a comparison of measured mean velocity and streamwise velocity variance from experiments and the LES. Similarly to the validation cases shown in previous sections, the MOSD-based WMLES results show good agreement with the turbulent statistics of the airflow in the presence of the waves, especially in the logarithmic region of the mean velocity profile. 
\begin{figure}
  \centerline{\includegraphics[width=12.4cm]{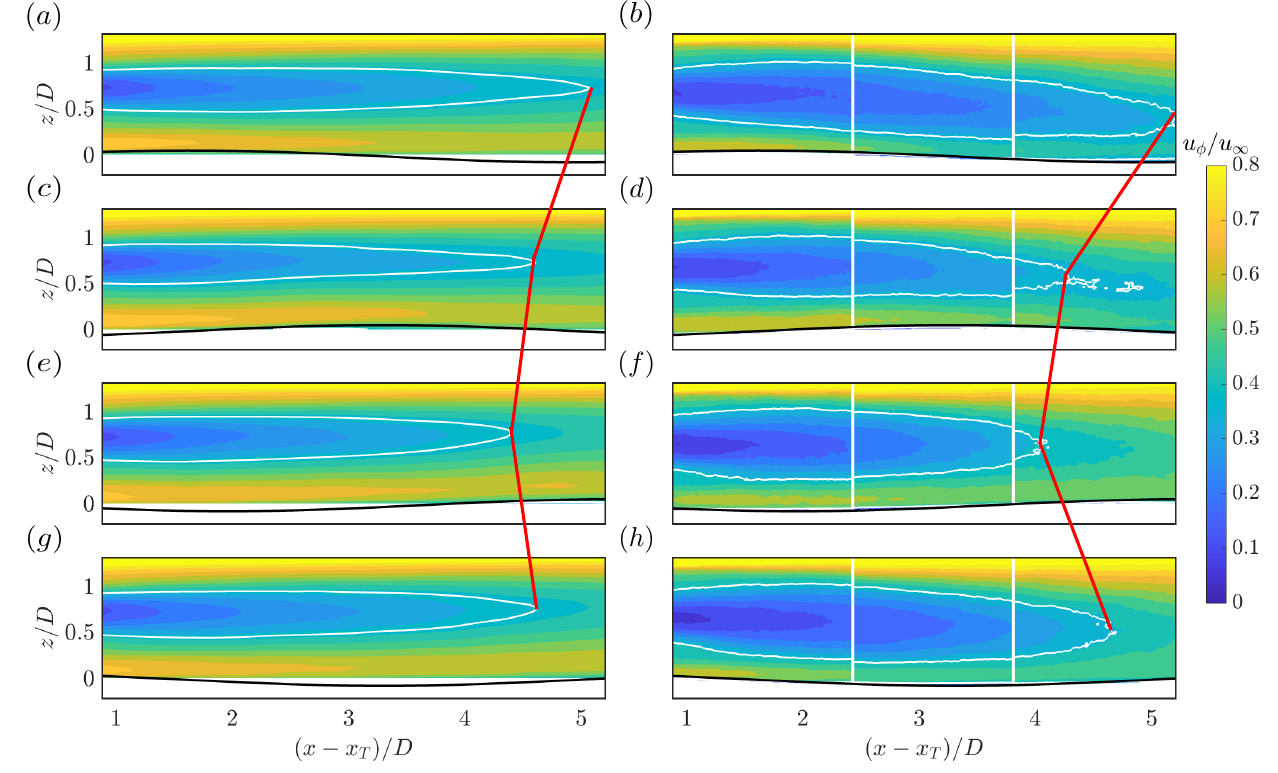}}
  \caption{Normalized phase-averaged streamwise velocity for MOSD-based WMLES (a,c,e,g), and experiments (b,d,f,h), respectively. (a,b) is for phase $\phi_1=\pi/2$,  (c,d) for $\phi_2=\pi$, (e,f) for $\phi_3=3\pi/2$, and (g,h) for $\phi_4=0,2\pi$. The red line indicates the phase-dependent wake length (where $u/u_\infty = 0.4$)}
\label{fig:14}
\end{figure}

To isolate the turbine wake motions specifically induced by the presence of surface waves, phase averaged streamwise velocity contours of the downstream velocity in the near and intermediate wake region are compared from the MOSD-based WMLES and the experiments. The phase averaging is performed at four distinct phases $(\phi_1,\phi_2,\phi_3,\phi_4) = (\pi/2, \pi, 3\pi/2, 2\pi$). For reference, $\phi = 0=2\pi$ is the phase in which the crest of the wave is located immediately under the turbine rotor location ($x_T$). Results comparing LES and experiments  in the near wake region starting one diameter downstream of the turbine up to about five diameters downstream are shown in Fig. \ref{fig:14}. 
The wake in the LES is slightly weaker and thinner than in the experiments and appears more horizontal than the experimental wake, which has a slight downward tilt. Differences due to the LES representation of the nacelle (which is significant in the wind tunnel setup) and in outer flow conditions (boundary layer in the experiments versus half-channel in LES) are expected and likely to contribute to observed differences in wake properties. However, in terms of trends with respect to the wave passage, there is good agreement. At each phase, we observe that in the some concavity in the lower regions of the wake  exactly where the crest of the wave is located, this behavior is best observed for $\phi_2 = \pi $. Elevated velocity regions near the wave at the bottom part of the turbine are also observed, these high velocity regions are elongated and follow the wave crest extending towards the trough of the wave. Interestingly, both the experimental and LES results indicate that as the wave passes underneath the wake downstream, the wake compresses and shortens in the streamwise direction. The wake is longest at phase $\phi_1$, e.g., consider the contour at a velocity of $40\%$ of the free-stream velocity highlighted with the white line. There the wave crest is near the turbine at $x/D \approx 1.5$. However, the wake then becomes shorter as the wave crest advances to $x/D \approx 3.3$ and beyond. The red line shown in Fig.~\ref{fig:14} for the WMLES marks the wake length at different phases for a fixed velocity threshold of $u/u_{infty}=0.4$. The compression and elongation of the length of the wake is primarily due to the pumping effect of the waves. As the wave moves, a horizontal velocity is induced onto the airflow above, increasing momentum and thus having an effect similar to wake recovery, or wake shortening at least for some of the time interval. This streching/compressing behavior of the wake can also be observed in the experimental data of \cite{fercak_2022} shown in the panels to the right in Fig. \ref{fig:14}. 
\begin{figure}
  \centerline{\includegraphics[width=12.4cm]{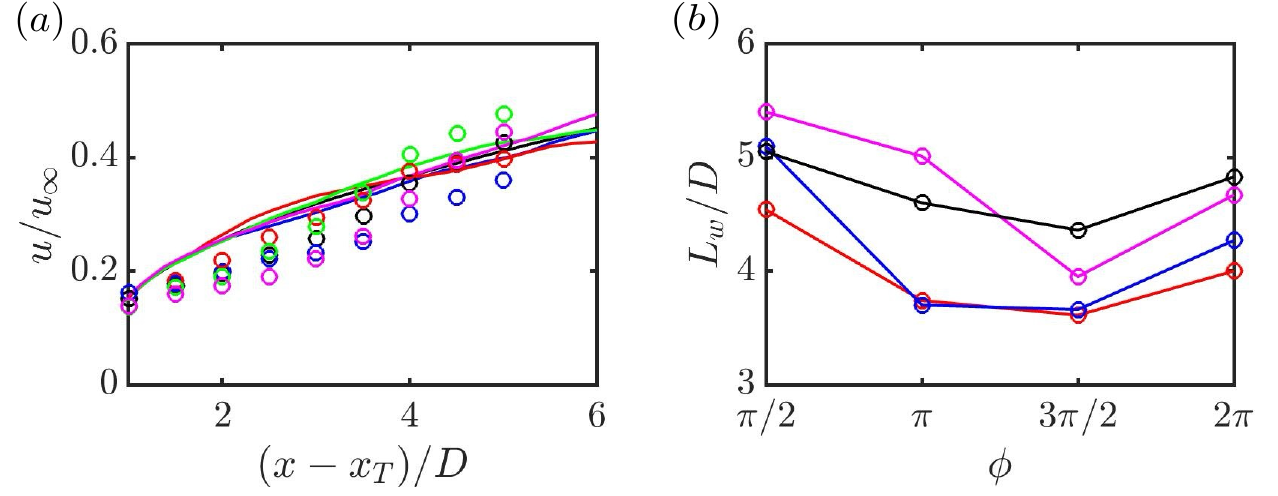}}
  \caption{ (a) Centerline velocity  at four phases. Solid lines: LES with MOSD wall model, circles: experiments from \cite{fercak_2022}. Blue: $\phi_1$, Red: $\phi_2$. Green: $\phi_3$, Pink: $\phi_4$.  (b) Wake length as function of phase. Red: $u/u_{\infty}=0.3$ (Exp), Blue: $u/u_{\infty}=0.35$ (Exp), Pink: $u/u_{\infty}=0.4$ (Exp), Black: $u/u_{\infty}=0.4$ (LES) }
\label{fig:15}
\end{figure}

For further comparison  between the numerical and experimental results, the wake velocity along the wake center as function of downstream distance from the turbine is calculated and shown in Fig.~\ref{fig:15}a. The wave steepness and velocity for this case can be  considered ``small'' and ``slow'', respectively, thus the vertical induced motions due to the wave movement are small but not negligible.  Finally, the wake length at each phase is shown in Fig.\ref{fig:15}b.  The experimental wake length was computed assuming three different threshold values of $u/u_{\infty}$, $0.3, 0.35$ and $0.4$. For the numerical results only the wake length at a threshold of $u/u_{\infty} = 0.4$ is shown, which is consistent with the behavior of the red line shown in Fig.~\ref{fig:14}. The maximum wave length is obtained at $\phi_1 = \pi/2$, while the minimum is at $\phi_1 = \pi$, for both the MOSD-based WMLES and the experimental data sets. After the wave crest advances from $x/D \approx 3.3$, the pumping effect is diminished and the wake size increases. This trend is consistent with the observations discussed before in Sect.\ref{sec:wave_indu}. Although there is good qualitative agreement, the observed remaining discrepancies between the WMLES phase-average results and the experiments are likely due to the  model's underestimation of the magnitude of  vertical motions, as discussed in Sect.\ref{sec:wave_indu}.
 
\section{Summary and Conclusions}
\label{sec:conclusions}
Understanding and modeling the physical processes and mechanisms within the marine atmospheric boundary layer hinges on the accurate quantification of momentum transfer at the air-sea interface. Presently, numerical methodologies employed to study fundamental aspects of wind-wave interactions typically entail computationally intensive techniques such as DNS or WRLES coupled with terrain-following grids or high-order spectral methods (``wave-phase-resolved'' approach) for resolving the wave. These approaches often incur prohibitively high computational costs and are difficult to implement in codes as simple boundary conditions, limiting their widespread applicability. Alternatively, effective roughness length scale-based methods (``wave-phase-averaged'' approach) can be imposed to characterize the wind-wave interactions in  LES. However, this methodology relies mostly on empirical parameters and does not enable the reproduction of any phase-dependent processes.  

In this study, a moving surface drag (MOSD) model for LES is introduced that effectively captures phase-dependent effects of waves, while maintaining computational efficiency. The proposed model focuses on accurately resolving the net pressure force resulting from the interaction between moving surface waves and turbulent wind. This model builds upon the surface-gradient based wall model introduced by \citep{AndersonMeneveau2010} that has also been extended to moving surfaces by \cite{aditya}.  The latter two approaches focus on the momentum flux of the air flow and assume that the entirety (or some modeled fraction) of the momentum is imparted onto the surface. In the present model, we extend the \cite{AndersonMeneveau2010} model along a different direction, namely by evaluating the potential flow pressure distribution on the windward facing side of the waves. Its integration yields the form drag (pressure force), which can be cast as a wall stress boundary condition.  Since the MOSD wall model is based on mechanistic principles, it does not in principle require further calibration or adjustments of empirical prefactors or adjustable parameters.

The model was validated against several experimental and numerical benchmark cases of turbulent boundary layer air flow over monochromatic moving waves. The MOSD model successfully captured the impact of waves on the mean velocity profile of the airflow across the majority of the test cases. However, while wave-induced structures show qualitative agreement with the comparison data, the strength of vertical motions is underpredicted by the model and future efforts should be focused on developing modifications to fully capture the induced vertical motions. To illustrate the applicability of the model to cases more complex than single wavelength and wind aligned waves, several such scenarios were simulated, demonstrating that the model effectively captures expected trends in such cases. We leave consideration of fully broadband wave fields for future work.

Finally,  a fixed-bottom offshore turbine was simulated in LES using the MOSD model to represent the waves and  results were compared to wind tunnel experimental data. Results confirm that features from copupled wind-wave-wake dynamics can be captured, specifically the stretching/compressing of the wake as the wave passes through. 

It is useful to underline the strong modeling assumptions that have been made in deriving the MOSD wall model. The modeling of unresolved roughness or the smooth wall viscous layer using the equilibrium model can be justified reasonably well.  While its underlying assumptions can be questioned (see e.g. the literature review and arguments presented in \cite{Fowler2022lagrangian}), it is a well-established methodology with many successful applications in geophysical and engineering flows. Conversely, the windward potential flow model is based on the pressure distribution of a potential ramp flow over a moving surface represents a strong idealization of the real flow conditions that can be expected near a moving wave. One important approximation is the use of potential flow. The real flow is turbulent and thus differences can be expected. However, possibly the strongest assumption is that in each LES cell that contains portions of the wave surface there exists an idealized ramp flow. In particular, we assume that the flow impinges horizontally upon the surface, separately and independently in each cell. Referring back to Fig. \ref{fig:potwave}(b), to the left of the cell shown there is in reality another cell where the flow could be deflected and arrive in a direction different than horizontal. To capture more realistically the near-surface flow even assuming potential flow would require taking into account non-local information, e.g., using a partial differential equation approach. For simplicity we purposefully limit the model to be entirely local and based only on quantities that are known at the cell where the wall model has to be evaluated. Under these constraints, the proposed model appears to be as accurate as can be expected. Its merits can be judged from the level of accuracy of the results presented.   While the initial application of MOSD wall model in LES show encouraging results, subsequent tests should be focused on applying the MOSD approach in LES of large offshore wind farms also including realistic ocean spectrum waves. And, while the model formulation is such that it can be applied to any multiscale time-dependent surface elevation, whether it provides accurate predictions in general cases and at larger surface slopes (e.g. approaching wave breaking conditions), remains to be seen.

\begin{acknowledgements}
 The support of the Percy Pierre Graduate Fellowship from Johns Hopkins University, and National Science Foundation grants \# CMMI-2034111, CMMI-2034160, CBET-2227263, and CBET-2037582 are gratefully acknowledged. 
\end{acknowledgements} 

\bibliographystyle{spbasic_updated}     
\bibliography{MOSDmodel-BLM}

\end{document}